\documentclass[conference,compsoc]{IEEEtran}

\usepackage{cite}
\usepackage{hyperref}

\usepackage{mathtools}
\usepackage[utf8]{inputenc}
\usepackage{booktabs}
\usepackage{xspace}         
\usepackage{graphicx}
\usepackage{pifont}
\usepackage{siunitx}
\usepackage{xcolor}
\usepackage{makecell}
\usepackage{multicol}
\usepackage{multirow}
\usepackage{subcaption}
\usepackage{bbm}
\usepackage{refcount}
\usepackage[capitalise]{cleveref}
\usepackage{array}
\usepackage{etoolbox}
\usepackage{totcount}

\newif\ifdraft
\drafttrue

\ifdraft
\newcommand{\at}[1]{\textcolor{cyan}{AT: #1}}
\newcommand{\cc}[1]{\textcolor{cyan}{CC: #1}}
\newcommand{\chris}[1]{\textcolor{cyan}{CC: #1}}
\newcommand{\ft}[1]{\textcolor{red}{FT: #1}}
\newcommand{\dpaleka}[1]{\textcolor{green}{DP: #1}}
\newcommand{\mcj}[1]{\textcolor{orange}{MCJ: #1}}
\newcommand{\hyrum}[1]{\textcolor{blue}{hyrum: #1}}
\newcommand{\kurt}[1]{\textcolor{purple}{KT: #1}}
\newcommand{\npc}[1]{\textcolor{purple}{NPC: #1}}

\else
\newcommand{\at}[1]{}
\newcommand{\cc}[1]{}
\newcommand{\chris}[1]{}
\newcommand{\ft}[1]{}
\newcommand{\dpaleka}[1]{}
\newcommand{\mcj}[1]{}
\newcommand{\hyrum}[1]{}
\newcommand{\kurt}[1]{}
\newcommand{\npc}[1]{}
\fi

\newcommand{\cmark}{\ding{51}}%
\newcommand{\xmark}{\ding{55}}%

\newcommand{\eg}{e.g., }

\renewcommand{\paragraph}[1]{\vspace{5pt}\noindent\textbf{#1.\xspace}}

\Crefname{thm}{Theorem}{Theorems}
\Crefname{cor}{Corollary}{Corollaries}
\Crefname{lem}{Lemma}{Lemmas}
\Crefname{prop}{Proposition}{Propositions}
\Crefname{assumption}{Assumption}{Assumptions}
\Crefname{definition}{Definition}{Definitions}
\Crefname{claim}{Claim}{Claims}
\Crefname{table}{Table}{Tables}
\Crefname{figure}{Figure}{Figures}

\newcounter{numdatasets}
\pretocmd{\tabular}{\setcounter{numdatasets}{0}}{}{} 
\newcolumntype{L}{>{\stepcounter{numdatasets}}l}
\regtotcounter{numdatasets}

\regtotcounter{numheaders}
\newcommand{\totaldatasets}{10}

\newcommand{\poisonrate}{0.01\%}%

\usepackage{threeparttable}

\newcommand{\snaptime}{t}
\newcommand{\snapstart}{\snaptime_0}
\newcommand{\prev}{\textrm{prev}}
\newcommand{\snapstartPrev}{\snaptime_{0, \prev}}
\newcommand{\snaptimeIth}{\snaptime_i}
\newcommand{\snaptimeIthPrev}{\snaptime_{i,\prev}}
\newcommand{\snaptimeIthLow}{\snaptimeIth^{\textrm{low}}}
\newcommand{\snaptimeIthHigh}{\snaptimeIth^{\textrm{high}}}
\newcommand{\snaptimeIthPrevLow}{\snaptimeIthPrev^{\textrm{low}}}
\newcommand{\snaptimeIthPrevHigh}{\snaptimeIthPrev^{\textrm{high}}}
\newcommand{\snaptimeIthPrevEstimate}{\hat{\snaptime}_{i,\prev}}
\newcommand{\snaptimeIthEstimate}{\hat{\snaptime}_i}

\begin{document}
%
\title{Poisoning Web-Scale Training Datasets is Practical}

\author{Nicholas Carlini$^1$ \quad Matthew Jagielski$^1$ \quad Christopher A. Choquette-Choo$^1$ \quad Daniel Paleka$^2$ \\
Will Pearce$^3$ \quad Hyrum Anderson$^4$ \quad Andreas Terzis$^1$ \quad Kurt Thomas$^5$ \quad Florian Tramèr$^2$ \\
\emph{$^1$Google DeepMind \qquad $^2$ETH Zurich \qquad $^3$NVIDIA \qquad $^4$Robust Intelligence \qquad $^5$Google}
}


%


\maketitle


\begin{abstract}
    Deep learning models are often trained on distributed, web-scale datasets crawled from the internet.
    In this paper, we introduce two new dataset \emph{poisoning attacks} that intentionally introduce malicious examples to a model's performance.
    Our attacks are immediately practical and could, today, poison $\totaldatasets{}$
    popular datasets.
    Our first attack, \emph{split-view poisoning}, exploits the mutable nature of internet content to ensure a dataset annotator's initial view of the dataset differs from the view downloaded by subsequent clients. 
    By exploiting specific invalid trust assumptions,
    we show how we could have poisoned $0.01\%$
    of the LAION-400M or COYO-700M datasets for just \$60 USD. 
    Our second attack, \emph{frontrunning poisoning}, targets web-scale datasets that periodically snapshot crowd-sourced content---such as Wikipedia---where an attacker only needs a time-limited window to inject malicious examples. 
    In light of both attacks, we notify the maintainers of each affected dataset and recommended several low-overhead defenses.
    

\end{abstract}


\section{Introduction}
Datasets used to train deep learning models have grown from thousands of carefully-curated examples~\cite{cifar10,deng2009imagenet,marcinkiewicz1994building} to
\emph{web-scale datasets} with billions of samples crawled from the internet~\cite{brown2020language,raffel2020exploring,schuhmann2022laion,whisper2022}.
At this scale, it is infeasible to manually curate and ensure the quality of each example. 
This quantity-over-quality tradeoff has so far been deemed acceptable,
both because modern neural networks are extremely resilient to large amounts of label noise~\cite{rolnick2017deep, zhang2021understanding}, and because training on noisy data can even improve model utility on out-of-distribution data~\cite{radford2021learning,radford2022robust}.

While large deep learning models are resilient to random noise, even minuscule amounts of \emph{adversarial} noise in training sets (i.e., a \emph{poisoning 
attack}~\cite{biggio2012poisoning}) suffices to introduce 
targeted mistakes in model behavior~\cite{carlini2021poisoning, carliniterzis2021poisoning, schuster2021you, wallace2020concealed}. 
These prior works argued that poisoning attacks on modern deep learning models are practical due to the lack of human curation. Yet, despite the potential threat, to our knowledge no real-world attacks involving poisoning of web-scale datasets have occurred.
One explanation is that prior research ignores the question of \emph{how} an adversary would ensure that their corrupted data would be incorporated into a web-scale dataset. 

Indeed
there is an \emph{exceptionally} vast literature \cite{biggio2011support,biggio2012poisoning,munoz2017towards,gu2017badnets,chen2017targeted,yao2019latent,schuster2021you,tramer2022truth,carlini2021poisoning,carliniterzis2021poisoning,li2021backdoor,wang2020attack,aghakhani2023venomave,ashcraft2021poisoning,barni2019new},
 \cite{saha2020hidden,turner2018clean,turner2019label,shafahi2018poison,doan2020februus,liu2019abs,shen2021backdoor,huang2020metapoison,qi2021hidden,souri2022sleeper,li2021invisible,zhong2020backdoor,geiping2020witches}
 \cite{biggio2011bagging,jagielski2018manipulating,xiao2015feature,chen2018detecting,tran2018spectral,xu2021detecting,gao2019strip,wu2021adversarial,shan2022poison,li2021anti,huang2019neuroninspect,guo2020towards,qiu2021deepsweep},
 \cite{salem2022dynamic,liu2020reflection,nguyen2020input,nguyen2021wanet,ge2021anti,quiring2020backdooring} that first presumes an adversary can modify a training dataset,
and then asks
(1) {what impact this could have},
(2) {if poisoning can be stealthy},
(3) {how to defend against poisoning},
and (4) {how to attack these defenses}.

Our paper does not address any of these questions as there are already
hundreds of papers already dedicated to each.
We focus on the preliminary question:
is it actually possible for an adversary to actually poison a dataset?

This paper introduces two novel poisoning attacks that \emph{guarantee} malicious examples will appear in web-scale datasets used for training the largest machine learning models in production today. Our attacks exploit critical weaknesses in the current trust assumptions of web-scale datasets: due to a combination of monetary, privacy, and legal restrictions, many existing datasets are not published as static, standalone artifacts. Instead, datasets either consist of an \emph{index} of web content that individual clients must crawl; or a periodic \emph{snapshot} of web content that clients download. This allows an attacker to know with certainty \emph{what} web content to poison (and even \emph{when} to poison this content).

%



%
%

Our two attacks work as follows:

\vspace{-3pt}
\begin{itemize}
\item \textbf{Split-view data poisoning:}
Our first attack targets current large datasets (e.g., LAION-400M) and exploits the fact that the data seen by the dataset curator at collection time might differ (significantly and arbitrarily) from the data seen by the end-user at training time. This attack is feasible due to a lack of (cryptographic) integrity protections: there is no guarantee that clients observe the same data when they crawl a page as when the dataset maintainer added it to the index.


\item 
\textbf{Frontrunning data poisoning:}
Our second attack exploits popular datasets that consists of periodical snapshots of user-generated content---e.g., Wikipedia snapshots. Here, if an attacker can precisely time malicious modifications just prior to a snapshot for inclusion in a web-scale dataset, they can \emph{front-run} the collection procedure. This attack is feasible due to predictable snapshot schedules, latency in content moderation, and  snapshot immutability: even if a content moderator detects and reverts malicious modifications after-the-fact, the attacker's malicious content will persist in the snapshot used for training deep learning models.


\end{itemize}

We explore the feasibility of both of these attacks in practice on 10 popular web-scale datasets. 
We show these attacks are practical and realistic even for a low-resourced attacker: for just \$60 USD, we could have poisoned 0.01\% of the LAION-400M or COYO-700M datasets in 2023.


To counteract these attacks, we propose two defenses:
\begin{itemize}
\item 
\textbf{Integrity verification} prevents split-view poisoning by distributing cryptographic hashes for all content, thus ensuring that clients observe the same data as when maintainers first indexed and annotated it. 

\item 
\textbf{Timing-based defenses} prevent frontrunning poisoning by either randomizing the order in which data is snapshotted and placed into web-scale datasets; or delaying content prior to its inclusion into a snapshot and applying reversions from trusted moderators. 
\end{itemize}

We discuss limitations of these defenses (\eg in the case of integrity checks, preventing \emph{benign} modifications such as
re-encoding, re-sizing, or cropping images) and more robust, future-looking solutions with fewer trust assumptions.



\paragraph{Responsible disclosure}
We disclosed our results to the maintainers of each of the 10 datasets appearing in this study. Six of these datasets now follow our recommended implementation for integrity checks. We provided patches to the most popular web-scale dataset downloader to support integrity checks. Finally, we notified Wikipedia about the frontrunning vulnerability in their data collection process.

\section{Background \& Related Work}
Our work builds on existing knowledge of the risk of poisoning large datasets, but focuses on the practical exploit vectors to launch such an attack. We outline why web-scale datasets have become of critical importance, known security risks of web-scale datasets, as well as auxiliary dataset quality issues that stem from ingesting uncurated data into models.

\paragraph{Towards uncurated datasets}
Deep learning is most effective when applied to large datasets \cite{brown2020language,kaplan2020scaling}.
But curating such datasets is expensive,
and the availability of training data has become a limiting factor for improving model utility.
For example, the scaling laws observed in the 
recent Chinchilla \cite{hoffmann2022training} language model indicate that training a compute-optimal
500 billion parameter model would require
11 \emph{trillion} tokens of training data---over $10\times$ more
data than is currently used to train models of this size~\cite{chowdhery2022palm}.
%
%
To drastically scale dataset sizes, it has become common to scrape data from a wider and wider range of untrusted and uncurated web sources.

\paragraph{Security risk of poisoning attacks}
Uncurated training datasets are prime targets for \emph{poisoning attacks} \cite{biggio2012poisoning}.
For example, an adversary could modify the training dataset (``poisoning'' it) 
so that some targeted example will be misclassified by models trained on this dataset.
Early poisoning attacks targeted fully-supervised classifiers \cite{gu2019badnets,chen2017targeted,shafahi2018poison} trained on curated datasets.
These attacks often aim to be ``stealthy'', by altering data points in a manner indiscernible to human annotators~\cite{turner2018clean}.
Attacks on uncurated datasets do not require this strong property. 
On these datasets, poisoning rates as low as 0.001\% have been shown effective~\cite{carlini2021poisoning, carliniterzis2021poisoning} at certain classes of poisoning attacks, e.g., targeted misclassification or planting model ``backdoors''~\cite{chen2017targeted, gu2019badnets}.
%

It is thus known that \emph{if} an adversary were somehow able to poison a fraction of a web-scale dataset, then they could cause significant harm. However, it is not well understood \emph{how} an adversary could
place poisoned samples in any training dataset \emph{without guessing beforehand which parts of the web will be collected}.
This paper answers that question.



%
%

\paragraph{Auxiliary risks related to data quality}
Spending time and effort to curate datasets has benefits besides security.
Possibly most important among these is that uncurated data has serious implications
for fairness, bias, and ethics \cite{bolukbasi2016man,radford2021learning,wang2022revise}.
For example, Birhane \emph{et al.} \cite{birhane2021multimodal} note that LAION-400M has ``troublesome and explicit images and text pairs of rape, pornography, malign stereotypes, racist and ethnic slurs, and other extremely problematic content''. Language datasets also contain similarly harmful ``hate speech and sexually explicit content, even after filtering'' \cite{luccioni2021s}.

Dataset curation is not a perfect solution to these problems. %
Birhane \emph{et al.} \cite{birhane2021multimodal} note, ``without careful contextual analysis,
filtering mechanisms are likely to censor and erase marginalized experiences''.
Any filtering approach that selectively removes some data sources over
others should carefully consider both the resulting security implications,
as well as more general data quality metrics.

\section{Threat Model \& Attack Scenarios}\label{sec:threat_model}
Before presenting the implementation details of our attacks, we introduce key terminology, our threat model, and the high-level intuition behind our two attacks.

\subsection{Terminology} \label{ssec:terminology}
Because it is infeasible to distribute web-scale datasets as
standalone artifacts (due to the dataset size or regulatory concerns), current training datasets fall into two categories.

In the first category, a \textit{maintainer} generates a set of $N$ tuples $\{(url_i, c_i)\}_{i=1}^N$ consisting of a resource identifier $url_i$ and auxiliary data $c_i$ (typically a label).
We let $t_i$ denote the time at which the $i$-th sample was originally collected.
Critically, the maintainer does not provide a snapshot of the data associated with $url_i$, due either to untenable storage costs \cite{chen2020vggsound,kemelmacher2016megaface,bansal2017s,bansal2017umdfaces}, privacy concerns \cite{torralba200880,cao2018vggface2}, or copyright limitations \cite{changpinyo2021conceptual}.
As such, we refer to these as \textit{distributed datasets}.
One example is the LAION-5B dataset \cite{schuhmann2022laion} which consists of five billion tuples of image URLs and corresponding text captions---corresponding to several hundred terabytes of data.

In the second category of datasets, a \textit{curator} produces a snapshot of a dataset $\{x_i\}_{i=1}^N$, where each sample $x_i$ is drawn from a set of URLs $\{url_i\}_{i=1}^N$ at time $t_i$, and then makes this snapshot publicly available. 
Because data served by these URLs changes over time, the curator will frequently (e.g., monthly) re-collect
a dataset snapshot so that users have an up-to-date view of the data.
We refer to these as \textit{centralized datasets}.
For example, both Wikipedia and Common Crawl regularly produce snapshots of their entire database. This simplifies access for people training large models, while also discouraging researchers from re-scraping the database directly.

Once one of these two types of datasets is published,
a \textit{client} (e.g., a researcher or applied practitioner) downloads a local copy of the training dataset $\mathcal{D}$, either by crawling each URL for decentralized datasets $\{(url_i, c_i)\}_{i=1}^N$ at a future time $t'_i > t_i$, or by downloading the centralized dataset $\{x_i\}_{i=1}^N$. In practice, clients often use a \textit{downloader} tool developed and maintained by a third party.

\subsection{Threat Model}
We assume the existence of a relatively unskilled, low-resource adversary who can tamper with the contents of a small number of URLs $\{url_i\}_{i=1}^N$ at some point in time $\hat{t}_i$, such that when a client or curator accesses resource $i$ at a future time $t_i' > \hat{t}_i$, they receive a modified (poisoned) dataset $\mathcal{D}' \neq \mathcal{D}$. 
The difference between the poisoned and intended datasets must be sufficiently large such that a model $f$ trained on $\mathcal{D'}$ will produce poisoned results for some desired input. We let $S_{adv} \subset \{url_i\}_{i=1}^N$ denote the set of URLs an adversary can modify. 

We assume the adversary has no specialized or insider knowledge about the curator, downloader, or maintainer---other than knowledge of the set of URLs $\{url_i\}_{i=1}^N$ used to generate $\mathcal{D}$ (since this information is published by the dataset curator).
We further assume all maintainers, curators and downloaders behave honestly and do not assist the adversary in any way. As such, the adversary has no control over the auxiliary data $c_i$ (e.g., supervised labels or text descriptions), nor can they add or remove any URLs from the training data that will be crawled by a client or curator. 

We make two critical (yet realistic) assumptions that enable our attacks.
For distributed datasets, we assume that clients do not compare the cryptographic \emph{integrity} of the local dataset copy $\mathcal{D'}$ that they downloaded, with the original dataset $\mathcal{D}$ indexed by the maintainer. For centralized datasets, we assume that it takes the curator at least some time $\Delta$ to detect malicious changes to the content hosted at any URL $url_i$ in the dataset (e.g., for Wikipedia, $\Delta$ is the time it takes to revert a malicious edit). Thus, the curator cannot detect that $url_i$ hosts poisoned content if the attacker poisoned the content at any time $t_i - \Delta \leq \hat{t}_i \leq t_i$, where $t_i$ is the time at which the content of $url_i$ is included in the dataset snapshot.
As we will show, these assumptions holds true for nearly \emph{all} modern web-scale datasets.
We discuss (in Section~\ref{sec:defenses}) how invalidating these assumptions---via cryptographic integrity checks and randomized crawling---can mitigate the attacks we describe.

\subsection{Attack Scenarios}
We propose two attack strategies for poisoning recent web-scale datasets.
In the subsequent sections we demonstrate the efficacy of these attacks in practice on real-world datasets, and describe the ethical safeguards we followed to minimize harm.
We focus our attacks on mechanisms that are unique to our study of dataset poisoning. 
Other potential security vulnerabilities
(e.g., the ability of an adversary to interfere with unencrypted network requests, or to exploit website vulnerabilities to inject new content) are out of scope and would only improve our attack success rates.

\paragraph{Split-view poisoning}
Our first attack exploits the fact that while the index of a distributed dataset published by a maintainer cannot be modified,
the content of URLs in the dataset can.\footnote{
Put differently, there is an important difference between the C types of ``\texttt{int const *}'' (how practitioners often treat these URL-based datasets) and ``\texttt{int * const}'' (what the dataset actually provides). }
This allows an adversary who can exert sustained control over a web resource to poison the resulting collected dataset collected by the end-user.

The specific vulnerability we exploit in our attack results from a fairly simple observation:
just because a web page hosted benign content when the dataset was
initially collected, this does not mean the same page is \emph{currently} hosting benign content.
In particular, domain names occasionally expire---and when they do, \emph{anyone can buy them}.
We show that domain name expiration is exceptionally common in large datasets. 
The adversary does not need to know the exact time at which clients will download the resource in the future: by owning the domain the adversary guarantees that \emph{any} future download will collect poisoned data.

We note that attackers already routinely buy expired domains to hijack the residual trust attached with these domains~\cite{lever2016domain,so2022domains, lauinger2018deletion}. 
Attackers have in the past targeted residual trust to
defunct banking domains~\cite{moore2014ghosts} and imported JavaScript libraries~\cite{nikiforakis2012you} to serve malware or steal user data,
to take over email addresses associated with the domain \cite{schlamp2015abandoned}, 
to control authoritative nameservers \cite{lever2016domain},
or simply to serve ads~\cite{lauinger2017expired}.
Here, we abuse residual trust to poison distributed datasets. While more sophisticated attacks may accomplish the same goal---such as exploiting a website, coercing a website's owner to modify content, or modifying unencrypted network traffic in flight---we focus on the natural phenomenon of domain expiration in this work.

To select domains to purchase, the adversary can either prioritize cheap domains that host multiple URLs in the dataset (minimizing the cost per poisoned URL), or domains that host content with specific auxiliary data $c_i$ (recall that the adversary \emph{cannot} modify the auxiliary data contained in the distributed dataset index). 
We show that split-view poisoning is effective in practice, as the index of most web-scale datasets remain unchanged long after their first publication, even after a significant fraction of the data goes stale. And critically, very few (and no modern) datasets include any form of cryptographic \emph{integrity} check of the downloaded content.

\paragraph{Frontrunning poisoning} 
Our second attack extends the scope of split-view poisoning to the setting where an adversary does \emph{not} have sustained control 
over web resources indexed by the dataset. Instead, the adversary can only modify web content for a short  period (e.g., a few minutes) before the malicious edits are detected.

This setting is common for datasets that aggregate content published on crowdsourced web pages, e.g., Wikipedia.
Indeed, it is easy to \emph{temporarily} edit Wikipedia to vandalize its contents \cite{stvilia2008information,shachaf2010beyond}.
A naive adversary might thus poison Wikipedia at arbitrary times and hope that 
a dataset curator will scrape the poisoned pages before the malicious edits are reverted.
However, Wikipedia vandalism is reverted in a few minutes on average~\cite{wiki:vandal},
so randomly-timed malicious edits are unlikely to affect dataset collection.

Our frontrunning attack relies on the fact that an adversary can, in some cases, predict \emph{exactly} when a web resource will be downloaded to the dataset.
As a result, an adversary can poison the dataset contents just prior to the curator's snapshot, thereby \emph{frontrunning} moderators might later revert the malicious edits. 
We will show that frontrunning attacks are particularly effective for Wikipedia datasets, because the official Wikipedia snapshot procedure accesses articles in a predictable---and well documented \cite{downloadthing}---linear sequence. An attacker can thus predict the snapshot time $t_i$ of any Wikipedia article down to the minute.

\section{Split-View Data Poisoning}\label{sec:retroactive-data-poisoning}

We begin our evaluation starting with
split-view data poisoning attacks,
where an attacker poisons a distributed dataset
by purchasing and modifying expired domains.

\subsection{Our Attack: Purchasing Expired Domains}
While split-view poisoning is applicable to any distributed dataset,
we focus on multimodal image-text datasets.
In such datasets, each URL points to an image hosted by some data provider,
and the auxiliary data contains a textual description of the image, e.g., an annotated class label or a caption extracted from the web page.

Our attack exploits the fact that the Domain Name System (DNS) does not assign
permanent ownership of any domain to a particular person or organization, 
but rather grants short ``leases'' that must be frequently renewed.
Thus, domain names continuously expire over time---intentionally or not---when the
re-registration fees are not paid.

When the domain that hosts an image in a distributed dataset expires,
\emph{anyone} can pay to take ownership 
over this domain and thereby gain the ability to return arbitrary content when the indexed image is later downloaded by a victim client.
Split-view poisoning abuses the residual trust inherent in an expired domain, as in traditional domain hijacking attacks~\cite{lever2016domain}.
We find that for many popular distributed datasets, domains are included with lax quality-assurance measures (if any), and thus domains with no special status that have been expired for months can freely be acquired to control a modest fraction of the entire dataset.

In this section we study whether it is possible to poison datasets by purchasing expired domains. We first quantify the fraction of domains that are expired in popular distributed datasets (\S\ref{sec:retroactivequantifying}),
then measure the frequency at which these datasets are scraped (\S\ref{sec:retroactiveimpact}),
verify our attack is not currently exploited in the wild (\S\ref{sec:inthewild})
and study the attack's potential down-stream impact (\S\ref{sec:retroactivetogether}).

\begin{table*}
\centering
\begin{threeparttable}[]
    \centering
    \caption{\textbf{All recently-published large datasets are vulnerable to \emph{split-view poisoning} attacks.} We have disclosed this vulnerability to the maintainers of affected datasets. All datasets have $>0.01\%$ of data purchaseable (in 2023), far exceeding the poisoning thresholds required in prior work~\cite{carlini2021poisoning}. Each of these datasets is regularly downloaded, with each download prior to our disclosure being vulnerable.}
    \vspace{-0.5em}
    \begin{tabular}{@{}L@{\hskip 0.38in}S[table-format=3.2]ccccr@{}}
    \toprule
        & \textbf{Size} & \textbf{Release} & \textbf{Cryptographic} 
        & \textbf{Data from} & \textbf{Data buyable} & \textbf{Downloads} \\%
        \textbf{Dataset name} & \textbf{($\times 10^6$)} & \textbf{date} & \textbf{hash?} 
        & \textbf{expired domains} & \textbf{for \$1K USD} & \textbf{per month} \\
        \midrule

        MMC4-FF~\cite{zhu2023multimodal} & 375 & \makebox[0.8em][l]{}2023\makebox[0.8em][l]{$^+$} & \xmark\makebox[0.8em][l]{}
        & 
        0.14\% & $\ge$ 0.01\% & -\\
        
        Falcon RefinedWeb~\cite{falconrefinedweb} & 276 & \makebox[0.8em][l]{}2023\makebox[0.8em][l]{$^+$} & \xmark\makebox[0.8em][l]{}
        & 
        0.24\% & $\ge$ 0.02\% & -\\
        
        OBELISC~\cite{laurencon2023obelisc} & 353 & \makebox[0.8em][l]{}2023\makebox[0.8em][l]{$^+$} & \xmark\makebox[0.8em][l]{}
        & 
        0.09\% & $\ge$ 0.01\% & -\\

        LAION-2B-en~\cite{schuhmann2022laion} & 2323 & \makebox[0.8em][l]{}2022\makebox[0.8em][l]{} & \xmark\makebox[0.8em][l]{$^\dag$}
        & 
        0.29\% & $\ge$ 0.01\% & $\ge$7\\
        
        LAION-2B-multi~\cite{schuhmann2022laion} & 2266 & \makebox[0.8em][l]{}2022\makebox[0.8em][l]{} & \xmark\makebox[0.8em][l]{$^\dag$} 
        & 
        0.55\% & $\ge$ 0.02\% & $\ge$4\\
        
        LAION-1B-nolang~\cite{schuhmann2022laion} & 1272 & \makebox[0.8em][l]{}2022\makebox[0.8em][l]{} & \xmark\makebox[0.8em][l]{$^\dag$}
        & 
        0.37\% & $\ge$ 0.02\% & $\ge$2\\
        
        COYO-700M~\cite{kakaobrain2022coyo-700m} & 747 & \makebox[0.8em][l]{}2022\makebox[0.8em][l]{} & \xmark\makebox[0.8em][l]{$^\ddag$}
        & 
        1.51\% & $\ge$ 0.10\% & $\ge$5 \\
        
        LAION-400M~\cite{schuhmann2021laion} & 408 & \makebox[0.8em][l]{}2021\makebox[0.8em][l]{} & \xmark\makebox[0.8em][l]{} & 
        0.71\% & $\ge$ 0.05\% & $\ge$10\\

        Conceptual 12M~\cite{changpinyo2021conceptual} & 12 & \makebox[0.8em][l]{}2021\makebox[0.8em][l]{} & \xmark\makebox[0.8em][l]{} & 
        1.19\% & $\ge $ 0.12\% & $\ge$33\\
        
        CC-3M~\cite{sharma2018conceptual} & 3.3 & \makebox[0.8em][l]{}2018\makebox[0.8em][l]{} & \xmark\makebox[0.8em][l]{} & 
        1.04\% & $\ge$ 0.08\% & $\ge$29\\
        
        VGG Face~\cite{parkhi2015deep} & 2.6 & \makebox[0.8em][l]{}2015\makebox[0.8em][l]{} & \xmark\makebox[0.8em][l]{} & 
        3.70\% & $\ge$ 0.17\% & $\ge$3\\
        
        FaceScrub~\cite{ng2014data} & 0.10 & \makebox[0.8em][l]{}2014\makebox[0.8em][l]{} & \cmark\makebox[0.75em][l]{$^\S$} &
        4.51\% & $\ge$ 0.61\% & $\ge$7\\
        
        PubFig~\cite{kumar2009attribute} & 0.06 & \makebox[0.8em][l]{}2010\makebox[0.8em][l]{} & \cmark\makebox[0.8em][l]{$^{\S*}$}
        & 
        6.48\% & $\ge$ 0.32\% & $\ge$15\\
        \bottomrule
    \end{tabular}
    \label{tab:pointer_datasets}
    
    \begin{tablenotes}
    \footnotesize
    \item[$+$] These datasets were published \emph{after} our paper was first released, and even still they do not provide cryptographic hashes.
    \item[$\dag$] LAION-5B released a ``joined'' dataset with hashes over the \emph{text} of the URL and Caption (not the \emph{contents} of the URL); this does not protect the integrity of the actual image.
    \item[$\ddag$] COYO-700M images are released with pHash \cite{phash} to validates benign image changes, but is not adversarially robust \cite{jain2022adversarial}.
    \item[$\S$] FaceScrub and PubFig contain cryptographic hashes, but the dataset maintainers do not provide an official downloader client that verifies these. We find that nearly all third-party downloaders for these datasets ignore hashes.
    \item[$*$] PubFig was initially released without hashes, but hashes were later added in Version 1.2 of the dataset.
  \end{tablenotes}
\end{threeparttable}
\end{table*}

\subsection{Quantifying the Attack Surface}
\label{sec:retroactivequantifying}
Table~\ref{tab:pointer_datasets} lists ten recent datasets we study in this paper that are 
vulnerable to split-view poisoning.
The three oldest datasets (PubFig, FaceScrub, and VGG Face) are datasets of faces
and associate each image with the identity of a single person (most often a popular celebrity).
The remaining seven datasets are multimodal datasets containing URLs that point to images, along with textual captions automatically extracted from the HTML of the webpage.
As such, for these datasets, the image can be modified by the owner of the corresponding domain, but the image's caption is fixed in the dataset index.

To measure the fraction of images that could be potentially poisoned,
we count the number of images hosted on domains that are \emph{expired} and \emph{buyable}.
We say that a domain is \textbf{expired} if the DNS record for that domain does not exist.
To measure this, we perform an \texttt{nslookup} on every domain name in the dataset from two geographically distinct data-centers in May 2022 and August 2022 and report the domain as expired if all four lookups result in a \texttt{NXDOMAIN} response.

\begin{figure}[htb]
    \vspace{-1em}
    \centering
    \includegraphics[scale=.72]{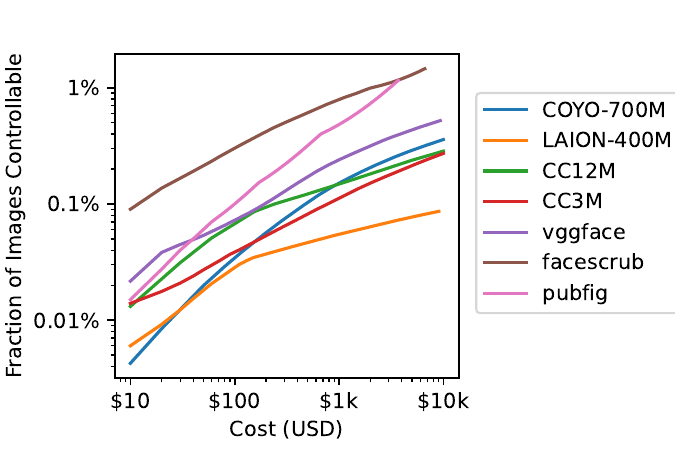}
    \vspace{-0.5em}
    \caption{\textbf{It often costs $\leq\$60$ USD to control at least $0.01\%$ of the data}. Costs are measured by purchasing domains in order of lowest cost per image first.}
    \label{fig:cost-poison}
\end{figure}

We further call a domain \textbf{buyable} if it is expired, and if at least one domain name registrar listed the domain as for sale by the registrar\footnote{%
Some registrars list domains as for sale even if they are actually owned by a squatter.
We exclude these domains from our set of buyable domains because purchasing these domains is often an expensive and lengthy process.} in August 2022.
Instead of counting the \emph{total} fraction of data that is buyable (which would represent a financially unconstrained adversary), Table~\ref{tab:pointer_datasets} reports the fraction of images in the dataset from domains that can be purchased for a total cost of \$1,000 USD. (Figure~\ref{fig:cost-poison} plots the fraction of datasets that can be purchased as a function of total cost.)
To compute this, we sort domains in decreasing order of the number of images the domain hosts divided by the cost to purchase this domain.

Overall, we see that an adversary with a modest budget could purchase control over at least $0.02\%$--$0.79\%$ of the images for each of the \totaldatasets{} datasets we study.
This is sufficient to launch existing poisoning attacks on uncurated datasets,
which often require poisoning just $\poisonrate$ of the data \cite{carliniterzis2021poisoning}.\footnote{Carlini \& Terzis~\cite{carliniterzis2021poisoning} assume the adversary can modify images \emph{and} their captions, whereas our adversary only controls images. In \S~\ref{sec:retroactivetogether}, we show modifying captions is unnecessary for a successful poisoning attack.}

There is also a direct relationship between the age of a dataset and how easy it is to poison.
Older datasets are more likely to contain expired domains, and therefore an adversary can purchase a larger fraction of the dataset.

\paragraph{Datasets are vulnerable from day zero}
A limitation of our above analysis is that we have measured the fraction of datasets vulnerable to poisoning in August 2022,
but many of these datasets were constructed years earlier.
It is therefore likely that many people who use these datasets would have already downloaded them at an earlier date,
and thus may not have been vulnerable to our poisoning attack
(although we will show in the following section that fresh downloads of these datasets remain frequent today).

Fortunately, the COYO-700M dataset was released during the writing of this research paper, on 30 August 2022.
On that same day, we computed the fraction of the dataset that was vulnerable to our poisoning attack and found that already $0.1\%$ of the images were hosted on expired domains that cost fewer than \$1,000 USD to purchase.
The reason that the number of expired domains is not zero on release is that
building these large datasets is a time-consuming and expensive procedure.
And so even though the COYO-700M index was released in August 2022, it took nearly a year to collect the dataset \cite{kakaobrain2022coyo-700m},
giving ample time for the earliest scraped domains to have expired before the dataset was released.

\paragraph{Measuring the attack cost} 
The most immediate question is if this attack can be realized in practice. 
The primary constraint in our attack is the monetary cost of purchasing domains. We measure this using the cost reported by Google Domains in July 2023. 
In \cref{fig:cost-poison} we show the fraction of images in a dataset that can be controlled by the attacker as a function of their budget. At least 0.01\% of each dataset can be controlled for less that $\$60$ USD per year. 

\subsection{Measuring the Attack Impact}
\label{sec:retroactiveimpact}

Our attacks are ``retroactive'' in the sense that we can poison a dataset after it has
been initially constructed by the curators---but the impact is limited to those who download it \emph{after} we take over the domains.
And given that it has been several years since many of these datasets were initially constructed,
it is not obvious that anyone would still download them by scrapping URLs instead of reusing a
previously downloaded version of the dataset.
As a result, in order to measure the potential impact of a split-view poisoning attack it is necessary to study the rate at which these datasets are actually still being actively downloaded by researchers and practitioners today.

\paragraph{Methodology}
We measure the rate at which each of these distributed datasets are downloaded
from the internet by purchasing multiple expired domains from each of the 10 listed datasets 
and passively monitoring requests to measure the rate at which URLs corresponding
to images from the distributed datasets are accessed.

For each dataset, 
we purchase the three most popular expired and buyable domains 
(that is, the domains available for purchase that hosted the most images),
and three randomly selected domains that were available for purchase.
We wrote a simple server to log all incoming HTTP
and HTTPS%
\footnote{To also monitor HTTPS traffic we obtained certificates from LetsEncrypt for each of our domains.}
requests, including the access time,
a hash of the IP address, the full URL being requested, and any additional headers provided.
This allowed us to count the frequency at which these datasets are still downloaded.
We ran this server for twelve months beginning in August 2022.

\paragraph{Analysis approach}
Our server received approximately $15$ million requests per month during our study,
a rate of 6 requests every second.
However from here it becomes necessary to separate the requests that were actually intending to download images from one of these datasets, from requests that come from other internet users or web crawlers.

To begin our analysis, we make a (significant) simplifying assumption that anyone downloading one of
these datasets does so from a single IP address.
We may thus underestimate the true rate at which each dataset is downloaded.
We then say that a particular IP address X downloads a dataset D at time $[T_0,T_1]$ if we meet both a precision
and recall requirement: 
\begin{enumerate}
    \item \textbf{Recall:} Within the time range $[T_0,T_1]$, the IP address X downloads at least $90\%$ of the URLs contained in $D$ under our control,
    including at least one URL from each of the 6 domains we own for that dataset.
    \item \textbf{Precision:} At least $50\%$ of the requests issued by X within this time range to the domains we control are to URLs in $D$.
\end{enumerate}

These conditions are conservative, but ensure we filter out web crawlers and other mass internet 
scrapers because these are likely to crawl other URLs from this domain (violating the precision 
constraint) or not crawl the majority of the URLs from the dataset (violating the recall constraint).
Because we own six domains per dataset it is exceptionally unlikely that, by random
chance alone, one particular IP will request URLs from each of these six otherwise-unrelated domains.
(In fact, we find that even checking 3 of these URLs would give almost identical precision.)
As a point of reference, for the CC-3M dataset, we received 51,000 image requests per month from a total of 2401 unique IPs.
By applying the precision constraint alone, we reduce this to 2007 unique IPs and 43,000 image requests;
by applying the recall constraint alone we get 70 unique IPs and 32,000 image requests;
and both together yields a further reduction of 64 unique IPs and 28,000 image requests (per month).

\subsubsection{Results}
We report our results in the rightmost column of~\cref{tab:pointer_datasets}. Even the oldest and least frequently accessed datasets still had at least 3 downloads per month. Thus, over the six months we tracked data, there were over 800 downloads that we could have poisoned with our attack.
Unsurprisingly, we found that newer datasets are requested more often than older datasets. Different datasets thus offer different tradeoffs for attackers: newer datasets have a smaller fraction of purchasable images, but an attack can reach many more vulnerable clients.

We observe that the largest billion-image datasets are downloaded significantly less often than smaller recent datasets.
We found that the reason for this is that 
these datasets are rarely downloaded in their entirety;
instead, they serve as an upstream source for smaller subsets.
For example, the Public Multimodal Dataset (PMD) \cite{singh2022flava}
and LAION-Aesthetics \cite{aesthetics} datasets consist almost
entirely of images drawn from LAION-2B-en.
Sub-datasets like this explain why sometimes we see IP addresses with very high precision but low recall.

\paragraph{Visualizing dataset crawlers}
Using our log files, we can visualize the ways in which dataset downloaders access these datasets by plotting URL requests as a function of time.
We order the set of URLs we bought for each dataset according to their order in the original dataset index.
In this way, crawlers that process the dataset index linearly should appear (roughly) as a linear line in our plot of URL requests over time.
To improve clarity, we assign each unique IP that accesses our server
a separate random color.

\begin{figure}[t]
    \centering
    \vspace{-1em}
    \includegraphics[width=\columnwidth]{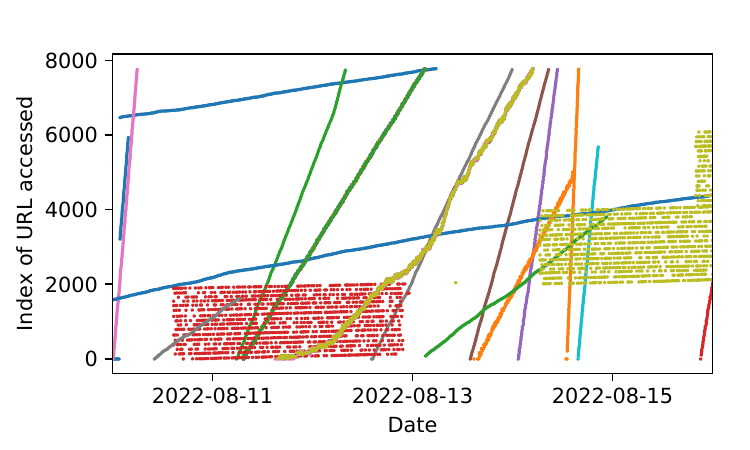}
    \vspace{-1.5em}
    \caption{Visualization of users downloading Conceptual 12M.
    By monitoring which URLs are requested from the domains we purchased,
    we plot every time a URL is requested over time, color coded by the source IP,
    and can directly read off several dozen users crawling Conceptual 12M.
    Figure~\ref{fig:overtime_raw} compares various filtering approaches.
    }
    \label{fig:overtime}
\end{figure}

Figure~\ref{fig:overtime} gives this plot for the Conceptual 12M dataset.
We find several trends in this data. First, most users who download this
dataset do so in a linear order from the first to the last URL.
However, the \emph{rate} at which URLs are accessed is highly variable: some downloaders crawl the entire dataset in a few hours, while others take several weeks to download the same dataset.
We also observe some users that batch the data into chunks and download each
chunk in parallel, as well as users that pause their download momentarily and then resume a few hours later (on a different IP).

While our strict precision and recall requirements already give strong evidence that the IP addresses we logged were indeed downloading the dataset, the linear ordering of URL requests from these addresses all but confirms this. Indeed, because the ordering of URLs in the dataset index is \emph{random} (instead of, say, alphabetical), 
a dataset download appears to be the only explanation for why the URLs would be linearly accessed in this particular order.

\paragraph{User-agent verification}
The most popular user agent%
\footnote{\texttt{Mozilla/5.0 (X11; Ubuntu; Linux x86\_64; rv:72.0) Gecko/20100101 Firefox/72.0}} is responsible for 77\% of the traffic to our domains.
This user agent is hardcoded\footnote{{\url{https://github.com/rom1504/img2dataset/blob/fc3fb2e/img2dataset/downloader.py\#L41}}}
in \texttt{img2dataset} tool \cite{beaumont-2021-img2dataset}, the most popular dataset crawler.
Given the claimed browser is Firefox 72--released in February 2020--it is highly likely that most of the requests indeed originate from \texttt{img2dataset}.

\paragraph{Ethical considerations}
We do not actually poison any datasets.
For all URLs we own, we return a 404 Not Found response; so 
from the perspective of a dataset downloader
our purchasing of the domain is completely transparent.
We also place a \texttt{robots.txt} file to prevent typical web-scrapers from
crawling our domains---this is unlikely to impact dataset downloads since dataset crawlers ignore this file.
Finally, a request to the root domain returns a 403 Forbidden with a response body explaining that this domain is part of a research study.
This response lists a contact email address
and offers to return the domain to the original
owner in case it was allowed to expire accidentally.
We have not received any contact on this address.
Appendix~\ref{sec:landing} contains the text of our landing page.

Our data collection is minimally invasive.
When searching for expired domains, we limit our
DNS requests to 500/second,
we only ask for the cost of purchasing the
top-10,000 domains in each dataset,
and only eventually purchase six.
This research study was deemed ``exempt'' by 
our institution's IRB.

\subsection{Is This Attack Exploited in the Wild?}
\label{sec:inthewild}

Given that this attack vector has existed for years against many datasets, and is easy to execute, it is not implausible that someone would have carried out such a poisoning attack in the past.
Yet, we could not find any evidence of this.

We search for a signature of a domain-purchasing attack by looking for domains that (1) host images that have been modified since the initial dataset release and (2) have changed ownership since the initial dataset release.
To detect image changes,
we can either check if images are perceptually similar to the originals (via, e.g., CLIP embeddings \cite{radford2019language}) or exactly identical (via a cryptographic hash).
Comparing images with cryptographic hashes has no false-negatives (i.e., any change is detected) but gives false-positives for ``benign'' changes, e.g., if a domain re-encodes or resizes its images.
In contrast, a perceptual hash has fewer false-positives but can have false-negatives if an adversary buys a domain and modifies images while preserving the perceptual hash (which is not collision-resistant).
Finally, to detect if a domain changed ownership, we request the \texttt{whois} record and check the last purchase date.

\paragraph{Results}
We perform our initial analysis on CC3M,
a dataset where we have the original raw images as ground truth.
Among all domains hosting more than 10 images, 
we find \emph{just one} domain has our attack signature when comparing images
with perceptual similarity.
Upon further investigation, we find that this domain has been purchased by
a domain squatter and any request to an image file on the domain returns an
advertisement.
If we instead compare cryptographic hashes of images, we find
two additional domains that have
been repurchased.
However, further investigation reveals the ownership of these domains
has not changed and the DNS record simply lapsed, and images were re-encoded.

We also conduct this same analysis on LAION-400M. 
Here we study three versions of the data: at original release (2021-11), and our own downloads at two later dates (2022-04 and 2022-08).
We find no domain with such a signature in LAION-400M, and thus have no evidence at present that this attack would have been exploited against this dataset.
Because we only have the original CLIP embeddings for this dataset (the original raw bytes were not saved), we only perform this comparison.
We find that by the first and second snapshot respectively, there are
$4.1$M and $4.2$M unique domains (of $5.6$M) that host at least one modified image---in total, we found $175$M and $183$M modified images, respectively. We measured this using a CLIP cosine similarity $<\!\!0.99$.
We randomly sample a few thousand domains including the $700$ with the most modified images.
We find many cases where domains are still owned by the original owner, are currently for sale, or have been redacted, but none appear malicious.

\subsection{Putting It All Together}
\label{sec:retroactivetogether}

Prior work \cite{carliniterzis2021poisoning} has successfully poisoned
multimodal models contrastively trained on datasets of 3 million images, 
under the assumption the adversary can \emph{arbitrarily} control the label of manipulated images.
However, in this paper we study datasets over $100\times$ larger,
and assume an adversary with \emph{no} control over the text captions.
Are poisoning attacks effective with these two changes?

We find they are.
We consider two poisoning attack objectives:
(1) cause a particular image to be misclassified as some target label from ImageNet,
and (2) cause a particular image to be classified as NSFW by the Stable Diffusion Safety Filter~\cite{rando2022red}.
For each of these attack objectives, 
we first identify appropriate text captions in LAION 400M for which the corresponding image domains can be purchased for a total of \$1,000 USD.
Then, we locally replace these images with poisoned samples to simulate the effect of an attack, without any potential to cause harm to others.

Specifically, we train one OpenCLIP~\cite{ilharco_gabriel_2021_5143773} model using a ViT-B-32 architecture for 32 epochs, at a batch size of 3072 on 16 A100 GPUs.
For our object-misclassification objective, we choose ten ImageNet classes that appeared in multiple text captions of images we can control.
When we poison 1,000 images (0.00025\% of the total dataset) our attack has a success rate of $60\%$ in flipping the model's zero-shot classification of the targeted image.
For our NSFW objective, we find captions corresponding to images (from buyable domains) labeled as UNSAFE in the LAION 400M dataset index.
Again at 1,000 poisoned samples, our attack has a success rate of above $90\%$.
More details about this experiment are in \Cref{apx:attack}.

\section{Frontrunning Poisoning}
Our second attack removes the assumption that the adversary has \emph{sustained} control over the web data in a training set.
To do this we make a new assumption: that we can predict precisely \emph{when} the web content will be downloaded. 
We will investigate this attack on Wikipedia-derived datasets, but also discuss how similar vulnerabilities may exist in the Common Crawl dataset in Appendix~\ref{app:vulnerable-text}.

\subsection{Our Attack: Editing Wikipedia}
\label{sec:frontrunning-wikipedia}
Wikipedia is a crowdsourced encyclopedia.
This makes it one of the most comprehensive and reliable datasets available on the internet~\cite{wiki:reliability}.
As a result of its quality and diversity, 
Wikipedia is frequently sourced for ML training data. Indeed, many language modelling datasets heavily rely on the English Wikipedia, e.g., it formed over 75\% of the words in the BERT training set~\cite{devlin-etal-2019-bert}, 1.5\% of the Pile dataset~\cite{gao2020pile}, and the entirety of the WikiText dataset~\cite{merity2016pointer}.
Many task-specific datasets also rely on the English Wikipedia: 
e.g., the WikiQA~\cite{yang2015wikiqa} question answering dataset (30,000+ downloads), and
the WikiBio~\cite{DBLP:journals/corr/LebretGA16} biography writing dataset (19,000+ downloads).
Finally, some of the distributed datasets discussed in \Cref{sec:retroactive-data-poisoning} index many images from Wikipedia articles.

Because Wikipedia is a \emph{live} resource that anyone can edit,
an attacker can poison a training set sourced from Wikipedia by making malicious edits.
Deliberate malicious edits (or ``vandalism'') are not uncommon
on Wikipedia, but are often manually reverted within a few minutes~\cite{wiki:vandal}.
As a result, actually poisoning Wikipedia appears challenging:
unlike the attacks in our prior section, an adversary cannot exert sustained control of any particular page and would thus have to hope that their malicious edit is timed just perfectly to affect a dataset download before being reverted.
 
However, we make one key observation that will guarantee the success of our poisoning attack: Wikipedia-derived datasets are not themselves live, but rather a collection of static snapshots.
This is because Wikipedia forbids using web crawlers to scrape the live website. %
Instead, Wikipedia makes available regular ``dumps'' (or snapshots) of the entire encyclopedia. 
Thus, training datasets sourced from Wikipedia use these snapshots instead of data crawled directly from the site.
For example, the authors of the BERT model~\cite{devlin-etal-2019-bert} explicitly recommend ``to download the latest [Wikipedia] dump'' to reproduce their results. 

This makes it possible to mount what we call a \textbf{frontrunning attack}.
An attacker who can predict \emph{when} a Wikipedia page will be scraped for inclusion in the snapshot can perform poisoning immediately prior to scraping.
Even if the edit is quickly reverted, the snapshot will contain the malicious content---\emph{forever}.
The attentive reader may argue we have not gained much: instead of having to predict the time at which Wikipedia is crawled by the end-user to produce a training set, the attacker has to predict the time at which Wikipedia is crawled to produce an official snapshot. But as we will see, the latter is actually easy.

This section explores how an adversary can time malicious edits to guarantee successful poisoning of a Wikipedia snapshot.
To this end, we need to answer two questions:
\begin{enumerate}
    \item How accurately can we predict when a page will be scraped for inclusion in a Wikipedia snapshot?
    \item How quickly do malicious edits get reverted?
\end{enumerate}

\subsection{Predicting Checkpoint Times}\label{ssec:preficting-checkpoint-times}

Wikipedia produces snapshots using a deterministic, well-documented protocol (with details that are easy to reverse through inspection). 
This makes it possible to predict snapshot times of individual articles with high accuracy.
\label{sec:enwikipred}

\begin{figure}
    \centering
    \includegraphics[scale=.5]{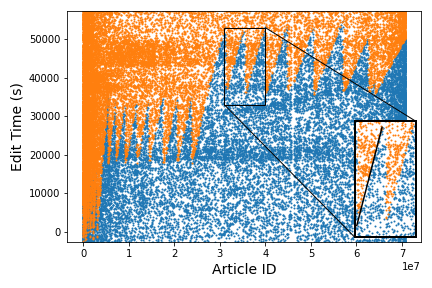}
    \vspace{-1em}
    \caption{\textbf{An adversary can easily predict when any given Wikipedia article will be
    snapshot for inclusion in the bimonthly dump.}
    We visualize edits around the June 1st, 2022 Wikipedia snapshot. 
    Each point corresponds to an edit made to a Wikipedia article, with the article ID on the X axis and time (in seconds) that the edit was made on the Y axis.
    Edit points colored \textcolor{blue}{blue} were \emph{included} in the snapshot, and edits colored \textcolor{orange}{orange} were \emph{not} included.
    The ``sawtooth'' pattern exhibited in the plot indicates a trend where multiple parallel jobs crawl Wikipedia articles sequentially
    to construct the snapshot. Furthermore, these parallel jobs run almost perfectly linearly through their allocated pages.}
    \label{fig:wikisaw}
\end{figure}

\subsubsection{How Wikipedia Snapshots Work}
English Wikipedia is archived on the 1st and 20th of each month.
The snapshot is produced by $n$ parallel workers; all Wikipedia articles are ordered sequentially by their ID and split into $n$ chunks, and each worker independently and linearly scrapes all articles in their chunk.

Due to Wikipedia's size, the whole process takes nearly a day to complete. Different articles thus get scraped at significantly different wall-clock times. As a result an edit at time $t_i$ for one article may be excluded from the snapshot, while an edit at time $t_j>t_i$ for a different article might be included.
Figure~\ref{fig:wikisaw} visualizes this ``sawtooth'' effect:
there are many edits (in blue) that are included in the June 1st dump that were made \emph{before} (i.e., below) a different edit that was \textbf{not} included (in orange). 

For a frontrunning attack to succeed, it is thus not sufficient to just predict the time at which the snapshot procedure begins. The attacker also needs to predict the precise time at which each individual page is scraped.

\subsubsection{Exploiting Rolling Snapshots}
To precisely predict each article's scrape time, the attacker can exploit consistencies in Wikipedia's snapshot process.

First, the adversary knows precisely when each dump starts, because Wikimedia \emph{explicitly makes this information available} by publishing live statistics on ongoing snapshots.\footnote{On \url{https://dumps.wikimedia.org/backup-index.html}}
%

Second, the \emph{rate} at which articles are crawled in a dump remains consistent across dumps, and can thus be approximated from prior dumps (interestingly, crawls tend to speed up slightly over time).

With these two pieces of information, the attacker can predict precisely when an article will be crawled.
For an article $i$, we denote the time at which it is crawled for the current snapshot as $\snaptimeIth$ and its crawl time in the previous snapshot as $\snaptimeIthPrev$.  
We denote the start time of the current and previous snapshots (as reported by Wikimedia) as $\snapstart$ and $\snapstartPrev$ respectively.
Due to our first observation above, the attacker knows $\snapstart$ and $\snapstartPrev$. Due to our second observation, we have that $\snaptimeIth - \snapstart \approx \snaptimeIthPrev - \snapstartPrev$. 
This allows us to estimate the snapshot time of the $i$-th article as $\snaptimeIth \approx \snapstart + (\snaptimeIthPrev - \snapstartPrev)$.
But calculating this requires knowledge of $\snaptimeIthPrev$---the time at which the $i$-th article was crawled in the previous snapshot. We now discuss how to retroactively estimate this.

\subsubsection{Determining the Article Snapshot Time}
Wikipedia snapshots do not explicitly list the snapshot time for each article.
But Wikipedia does give some auxiliary information: a complete list of edits with the precise time every edit is made.
We show how this information can be used to retroactively estimate an article's snapshot time.

\begin{figure}
    \centering
    \includegraphics[width=\linewidth]{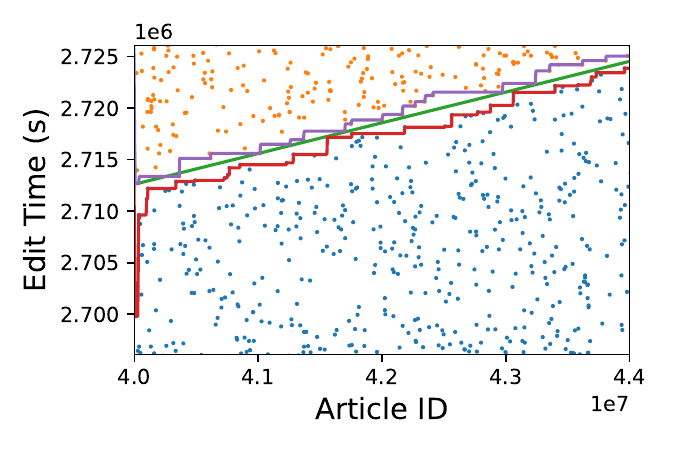}
    \caption{We can obtain tight estimates on the time at which each article is snapshot. The green and orange lines show the interval $[\snaptimeIthPrevLow, \snaptimeIthPrevHigh]$ for a range of articles from the English Wikipedia. On average, our predictions (blue line) are 27 minutes from the furthest interval boundary.}
    \label{fig:interval}
\end{figure}
Recall from Figure~\ref{fig:wikisaw} that for each article we can find the list of edits that were  included in the current snapshot (blue points), with later edits appearing in the next snapshot (orange points).
For each article, we thus know that the snapshot time $\snaptimeIth$ was in-between the times of the article's last included edit (top-most blue point) and the first non-included edit (bottom-most orange point).
But this interval is loose: the time between these edits is often several days.

To refine our estimate of the snapshot time for each article, we can again exploit the consistency present in the Wikipedia crawling process. We observe that articles are processed sequentially: by zooming in to just a single crawling job as shown in Figure~\ref{fig:wikisaw}, we see that articles are crawled sequentially, and a clear line separates the last-included and first-not-included edits of each article. That is, for articles $i$ and $j$ processed in the same job, we have that $\snaptimeIth < \snaptime_j$ if $i < j$. We can thus tighten our interval around each article's edit time by continually tracking the most recent edit made before the snapshot (for each article, this is the top-most blue edit made on an earlier article in that job), as well as the earliest edit among all subsequent articles in the job that was not included in the snapshot.
We visualize this in Figure~\ref{fig:interval}. 
For each article in the previous snapshot, we can thus obtain a time interval $[\snaptimeIthPrevLow, \snaptimeIthPrevHigh]$ that contains the true (but unknown) snapshot time $\snaptimeIthPrev$. By our construction outlined above, we guarantee that the lower and upper limits of these intervals monotonically increase for all articles in a job (see \Cref{fig:interval}).

To produce an estimate $\snaptimeIthPrevEstimate$ for each article's previous snapshot time, we compute a  linear fit for the snapshot intervals of all articles processed by a single thread, as illustrated in \Cref{fig:interval}. This lets us predict the article's next snapshot time as $\snaptimeIthEstimate \approx \snapstart + (\snaptimeIthPrevEstimate - \snapstartPrev)$.


\begin{figure}
    \centering
    \includegraphics[width=\linewidth]{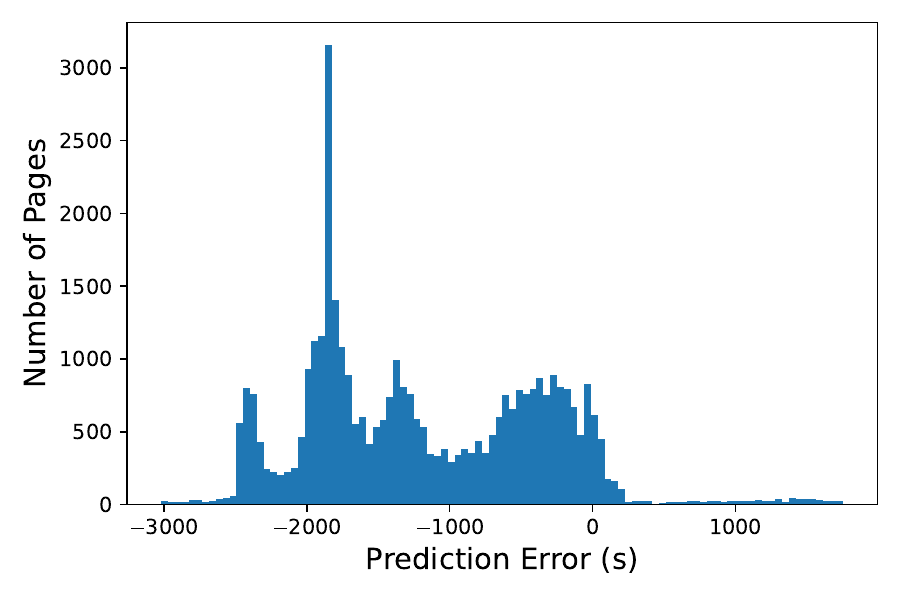}
    \vspace{-1em}
    \caption{Distribution of Wikipedia checkpoint time prediction errors. Most predicted checkpoint times are within 30 minutes of our constructed ground truth. In general, we predict edits too early, so it is important to later adjust for this bias, as we will discuss in Section~\ref{sec:wikioverall}.}
    \label{fig:pred_errs_wiki_en}
\end{figure}

\subsubsection{Evaluating our Predictions}
\label{ssec:wiki_eval_preds}
We now evaluate our procedure for estimating article snapshot times.
Ideally, we would directly compare our predicted snapshot times $\snaptimeIthEstimate$ with the true snapshot time of the $i$-th article. But we do not know the ground truth, except up to some interval 
$[\snaptimeIthLow, \snaptimeIthHigh]$ which we can compute a posteriori as described above. 
We thus proceed in two steps.

First, we show that our linear fit to estimate the previous snapshot time $\snaptimeIthPrevEstimate$ is accurate.
For this we measure the maximum absolute error between the predicted time $\snaptimeIthPrevEstimate$, and the unknown ground truth in the interval $[\snaptimeIthPrevLow, \snaptimeIthPrevHigh]$. This provides an upper bound on the true estimation error---bounded by 27 minutes on average.
%


Second, we evaluate the accuracy of the \emph{extrapolation} of our predictions from one snapshot to the next. That is, we evaluate how close our \emph{a priori} predicted snapshot time $\snaptimeIthEstimate \coloneqq \snapstart + (\snaptimeIthPrevEstimate - \snapstartPrev)$ is to the snapshot time that we could have estimated \emph{a posteriori} using the linear fit described above. \Cref{fig:pred_errs_wiki_en} shows the distribution of the error estimates. Our predictions are correct to within roughly 30 minutes in most cases. We notice, however, that our extrapolation errors are biased towards negative. We find that this is because snapshots slightly speed up over time, so we typically overestimate the next snapshot time of an article. We will correct for this in \Cref{sec:wikioverall}, when we produce a conservative estimate of our attack's success in poisoning Wikipedia snapshots.

\subsection{Estimating Revision Speed}
\label{sec:revspeeden}
Now that we have measured how accurately we can predict when a future snapshot will happen, we turn to measuring the size of the opportunity window to make a malicious edit before it is reverted.

Note that while the most accurate methodology would be to inject malicious edits and measure the distribution of reversion times, we believe this would be unethical. Instead, we take an entirely \emph{passive}---albeit less accurate---approach as discussed in Section~\ref{ssec:wiki-ethics}.

\begin{figure}[t]
    \centering
    \includegraphics[width=\linewidth]{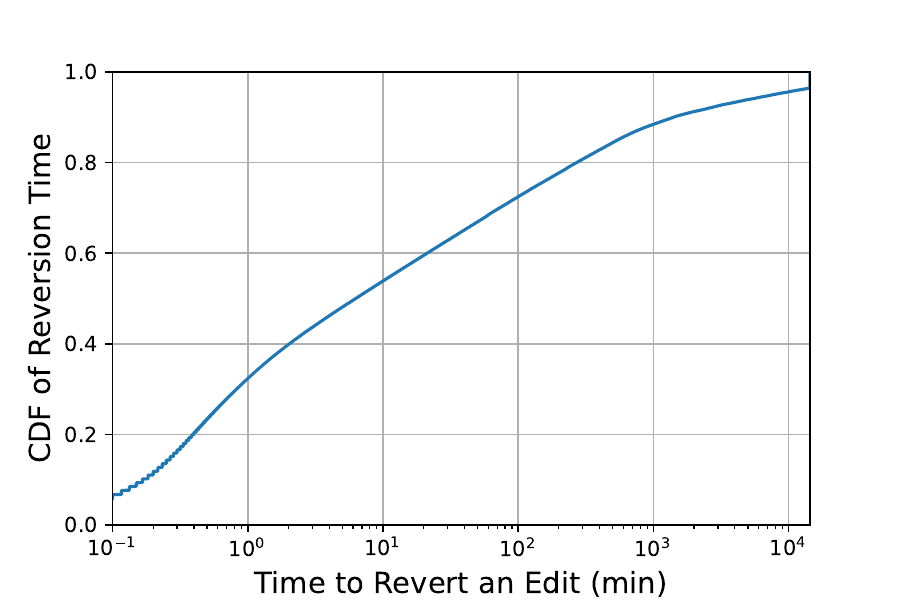}
    \vspace{-1em}
    \caption{A CDF of revision times for English Wikipedia. Roughly 35\% of revisions take more than 30 minutes.}
    \label{fig:revtime_wiki_en}
\end{figure}

To measure the speed of revisions, we construct a dataset of all edits made to Wikipedia from January 2021 to June 2022 (for a total of 18 months), and classify every edit as either an addition or as a reversion if they contain one of a fixed set of strings\footnote{This set of strings is produced by manual analysis of a sample of comments from each Wikipedia; details are given in Appendix \ref{app:annotating}.} which are frequently used in reversion comments. 
We then \emph{conservatively} assume that the edit being reverted was the immediately preceding edit, and so measure the reversion time as the elapsed interval between these two edits.\footnote{This under-reports the edit time because if the vandalism was from an earlier edit, we would incorrectly use the later edit's time instead.}
Figure~\ref{fig:revtime_wiki_en} plots this distribution. 
When we combine the roughly 30 minutes of error in predicting future snapshot times (c.f.~\Cref{fig:pred_errs_wiki_en}), with another roughly 30 minutes for the average uncertainty in our estimate of the true snapshot time (c.f.~\Cref{fig:interval}), we can conservatively estimate that the attacker can time their edit so as to be within one hour, on average, of the true snapshot time.
Approximately $32\%$ of reversions take more than an to be reverted and so the attack is likely to succeed often. In the next section, we refine this estimate to determine more precisely how many articles we could have poisoned.

%

\subsection{Putting It All Together}
\label{sec:wikioverall}
Using our predictions of relative article snapshot times, our interval bound on the true snapshot time, and the distribution of reversion times, we can (conservatively) estimate the fraction of Wikipedia an adversary could have poisoned.
There are two potential ``failure cases'' where a malicious edit 
may not make it into the dump: 
\begin{itemize}
    \item the malicious edit is applied too late: the article was already snapshot, or
    \item the malicious edit is applied too early: the edit gets reverted before the article is snapshot. 
\end{itemize}
This induces a tradeoff: the attacker should make edits early enough to ensure they do not miss the snapshot time, but late enough to maximize the chance of frontrunning editors.

We therefore compute the optimal time to apply a malicious edit as follows.
Recall
that we use $[\snaptimeIthLow, \snaptimeIthHigh]$ to represent the tightest interval around the true (but unknown) snapshot time $\snaptimeIth$ of the $i$-th article, and $\snaptimeIthEstimate$ is the predicted snapshot time.
To balance the two failure modes above, and to account for the bias in our predictions (see \Cref{ssec:wiki_eval_preds}), we introduce an ``adjustment'' variable $a$ so that the adversary adds their malicious edits at time $\snaptimeIthEstimate + a$ instead of exactly at time $\snaptimeIthEstimate$.

Then the fraction of malicious Wikipedia edits that will make it into the snapshot, when they are made at time $\snaptimeIthEstimate + a$, can be lower-bounded as:
\[
\mathcal{A}(a) = \frac{1}{|D|}\sum_{i\in D}
(1 - \overbrace{p_{rev}(\snaptimeIthEstimate + a ; \snaptimeIthHigh)}^{\text{Edit applied too early}})
\cdot
(1-\underbrace{\mathbbm{1}[\snaptimeIthEstimate + a> \snaptimeIthLow]}_{\text{Edit applied too late}}) \;,
\]
where $\mathbbm{1}[\snaptimeIthEstimate + a> \snaptimeIthLow]$ is the 
indicator function that is one if the edit is applied after the checkpoint (here we conservatively use our lower bound $\snaptimeIthLow$ on the true checkpoint time),
and $p_{rev}(\snaptimeIthEstimate + a ; \snaptimeIthHigh)$ is the probability that the edit is reverted before the checkpoint (here we conservatively use the upper bound $\snaptimeIthHigh$ on the true checkpoint time).

We compute this sum using our results from Section~\ref{sec:enwikipred} and Section~\ref{sec:revspeeden}. By taking the maximum over a sweep of potential $a$ values, we obtain $\text{max}_a\, A(a)=0.065$. According to our conservative analysis, \textbf{we can poison 6.5\% of Wikipedia documents} absent any other defensive measures. 


In reality, of course, a number of factors beyond our analysis would likely prevent us from reaching this fraction, such as rate limiting of edits or IP bans.
We also ``cheat'' in choosing the optimal value of the adjustment value $a$, but we do not consider this a major limitation---an adversary could likely use more historical data to produce better estimates $\snaptimeIthEstimate$ as well as good estimates of $a$. However, our analysis is also pessimistic since we assume we only try \emph{once} to poison any given article. An adversary could attempt a more targeted attack 
where they retry edits on targeted articles, to force editors to revert multiple times and increase the likelihood of the edit making it into the dump. Ultimately, our best-effort estimate of 6.5\% of poisoning success is orders-of-magnitude higher than what is required in prior poisoning attacks~\cite{carliniterzis2021poisoning}. We thus argue that a successful frontrunning poisoning attack on Wikipedia snapshots is practical, and that finding ways to mitigate such attacks is a worthwhile research direction.

\subsection{Multilingual Wikipedia}
Wikipedia is also frequently used for non-English language modeling.
For example, the multilingual version of BERT is trained entirely on the top 104 Wikipedia languages.
Multilingual datasets often rely \emph{more heavily} on Wikipedia than English datasets. Thus, poisoning Wikipedia is even more harmful for non-English language modeling tasks.
To measure this vulnerability, we investigate the Wiki-40B dataset~\cite{guo2020wiki} which is frequently used to train large multilingual models.

We repeat our analysis from the previous section on 35 of the 39 non-English languages contained in Wiki-40B by
identifying which strings often represent a reversion in these languages.\footnote{We were unable to access the German, Chinese, and Czech checkpoints for the checkpoint times we studied, and Tagalog did not have enough data to reliably analyze.} 
Again our analysis here is loose: we identify only a subset of (often automated) reversions;
however for the same reason as above we believe this represents
a lower bound of the mean reversion time.

\begin{figure}
    \centering
    \includegraphics[width=\linewidth]{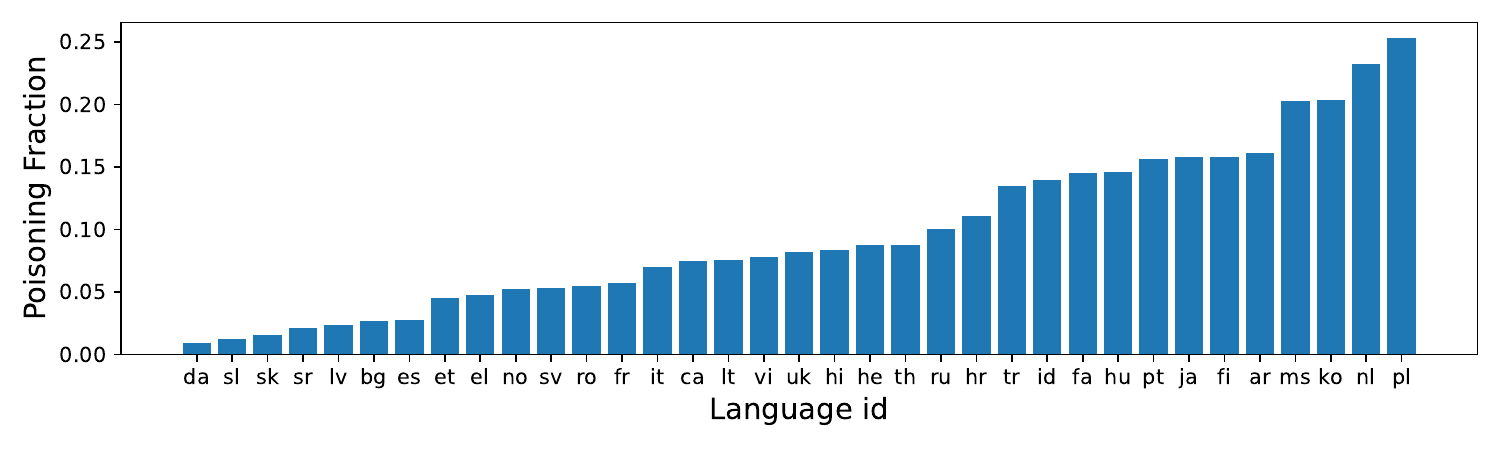}
    \vspace{-1.5em}
    \caption{\textbf{Multilingual Wikipedia is more vulnerable to frontrunning poisoning attacks.} We compute poisoning rates for 36 of the languages languages from Wiki-40B~\cite{guo2020wiki} by reusing our attack from Sections~\ref{ssec:preficting-checkpoint-times} to~\ref{sec:wikioverall}.}
    \label{fig:wiki_multiling}
\end{figure}

We find that 22 (63\%) of the non-English Wikipedias were easier to poison than the English Wikipedia, as shown in \cref{fig:wiki_multiling}. 
Feasible poisoning rates range from 0.95\% to as much as 25.3\%, with a median value of 8.2\%.
In general, the increase in vulnerability comes from multilingual Wikipedias having more predictable checkpoints, for two reasons. First, because these Wikipedias are smaller, the entire checkpointing procedure is shorter, reducing the amount of variance in checkpoint time between different pages. Second, because the Wikipedias change less between successive checkpoints, the speed of checkpointing is more stable, improving our predictions. This may be why some of the larger Wikipedias, such as Spanish, Danish, and Italian, have comparable poisoning rates to English Wikipedia. However, our interval-based measurement is also more conservative for languages with slower edits, as the intervals will be larger, giving very small lower bounds for some small Wikipedias, such as Slovak and Slovenian.

We reiterate that such large poisoning rates are unlikely to ever happen, due to IP bans or rate limiting. The most important takeaways from our analysis here are that 1) multilingual Wikipedias are vulnerable to poisoning, and often more vulnerable than English Wikipedia, and 2) multilingual datasets tend to rely more on Wikipedia than English datasets do, compounding this risk.

\subsection{Ethical Considerations}\label{ssec:wiki-ethics}

Our actions here are entirely passive.
We make no edits to Wikipedia and, aside from downloading datasets from the
official sources, never interact with Wikipedia.
While this leads to limitations in our analysis, we believe this is the correct way to run such a study to avoid harming the Wikipedia editor community.
We disclosed our attack analysis (and later defense recommendations) to researchers at Wikimedia who acknowledged the vulnerability
before release of this paper.

\section{Defenses}\label{sec:defenses}
In order to address the attacks that we identified, we propose an integrity-based defense for split-view poisoning and a timing-based defense for frontrunning poisoning. We also discuss potential directions to address poisoning more generally. We shared these defenses with curators, maintainers, and downloaders as part of our responsible disclosure and report on the status of their implementation of defenses.


%



\subsection{Existing Trust Assumptions}
Per our threat model (Section~\ref{sec:threat_model}), our proposed defenses assume that all maintainers, curators, and downloaders are trusted and behave honestly. This means that maintainers provide the same distributed dataset index$\{(url_i, c_i)\}_{i=1}^N$ to any client and that the index itself has not been poisoned (e.g., due to insider risk or compromise). Downloaders for distributed datasets honestly access each $url_i$ and compute integrity checks added via our defenses. Curators provide the same centralized dataset $\mathcal{D}$ to all clients, with curators controlling the time $t_i$ at which any element $url_i$ is snapshot. These assumptions mirror the existing trust that clients place in maintainers, curators, and downloaders and thus represent the quickest, short-term path to enacting defenses. We discuss limitations of these trust assumptions (and robust solutions with fewer trust assumptions) in Section~\ref{sec:defense_limitations}.

\subsection{Preventing Split-View Poisoning}\label{ssec:static-distributd-defenses}
While the simplest defense to split-view poisonings attack would be to convert the distributed dataset $\{(url_i, c_i)\}_{i=1}^N$ into a centralized dataset (\eg as in YFCC100M~\cite{thomee2016yfcc100m}), this is presently unrealistic due to the monetary, privacy, and legal challenges laid out in \cref{sec:threat_model}. Instead, maintainers---or another trusted third-party---can prevent split-view attacks by attaching a cryptographic hash $h_i=H(x_i)$ of the raw data $x_i$ obtained from $url_i$ at time $t_i$ prior to any attack. 
A downloader would then check whether $H(x'_i) = h_i$, where $x'_i$ is the content of $url_i$ at time $t'_i$. The downloader discards any data where the client and maintainer receive distinct content. Here, $H$ should be a cryptographic hash function such as SHA-256.

\paragraph{Implementation \& Responsible Disclosure}
Enacting this defense requires a number of ecosystem changes. Currently, only PubFig and FaceScrub include cryptographic hashes as part of their distributed dataset (see Table~\ref{tab:pointer_datasets}). 
And because these two datasets do not provide an official downloader, the community has relied on a number of third-party downloader scripts that for the most part do not actually verify these hashes.\footnote{We examine the 6 most popular downloader scripts for each dataset (gathered by searching for \texttt{``[pubfig|facescrub] dataset download github''}), and find that only one script per dataset implements hash verification. The one for FaceScrub
checks hashes by default, while the one for PubFig
requires users to run a separate verification script.}
Fortunately, for larger datasets the \texttt{img2dataset} \cite{beaumont-2021-img2dataset} tool has become the canonical
downloader used in $75\%$ of requests to our domains.

As part of our responsible disclosure, we reached out to every maintainer (see Table~\ref{tab:pointer_datasets}), suggesting the addition of SHA-256 hashes as part of the dataset index.
At the time of writing, CC3M, CC12M, LAION-2B-en, LAION-2b-multi, LAION-1B-nolang, and LAION-400M now release their dataset with SHA-256 hashes of the image contents.
We additionally implemented an option in \texttt{img2dataset} to verify SHA-256 image hashes upon download, thus preventing our attack for anyone using this tool,
and provide our own backup of hashes in a Google Cloud Bucket at
\texttt{gs://gresearch/distributed-dataset-hashes/}
for the datasets where we have (near-)original data.

\paragraph{Limitations}
Integrity checks are viable if benign content remains static.
If content is altered in any way (\eg by re-encoding, cropping, re-sizing, or uploading a higher-resolution image) the original hashes will no longer match. 
This can significantly degrade the utility of a dataset: for example, we obtain the original raw data from the first Conceptual Captions 3M dataset downloaded in 2018 and compare this to our most recent download of these same URLs in 2023.
Of the 3.3 million original images, 2.9 million images are still hosted online, but just 1.1 million  of these images have hashes that match the original---the other 1.8 million images have changed since the initial dataset.

This suggests that our defense, while providing perfect protection against split-view poisoning attacks,
has the potential to degrade utility.
%
Switching to a perceptual hash function (which aims for invariance to small image changes) would lead to higher utility, but would not meaningfully prevent our poisoning attacks because an attacker could upload poisoned images that were adversarially modified to fool the perceptual hash~\cite{hao2021s, struppek2022learning, jain2022adversarial}. 
This suggests qualitatively new defense ideas will be necessary to defend against our attacks without a high utility cost.

\subsection{Preventing Frontrunning Poisoning}\label{ssec:defenses-frontrunning}
Our frontrunning poisoning attack relies on the fact that an adversary
only needs sustained control of data for a few minutes to succeed.
To defend against this attack, it suffices to increase the duration $d = t_i - \hat{t}_i$ that an attacker must retain control over $url_i$ for it to be included in the snapshot at time $t_i$, where $\hat{t}_i$ indicates the time an attacker first modifies the URL's contents. 
If a curator can detect malicious modifications within time $\Delta$, then increasing $d > \Delta$ effectively thwarts the attack. 
This can be achieved in one of two ways: (1) curators can randomize the snapshot order of the $url_i$ and extend the time required to snapshot the complete corpus;
or (2) curators could freeze edits to the content of $url_i$ at time $t_i$, wait for a period $T > \Delta$ for edits to go through review, and then finally release the snapshot at time $t_i + T$.

\paragraph{Implementation \& Responsible Disclosure} 
For our first approach, Wikipedia could randomize its snapshot order of articles instead of its current sequential method based on article IDs. This thwarts an adversary's ability to predict precisely when an article will be selected for snapshotting, requiring they sustain control of articles for the entirety of the snapshot time, $t_n - t_0$, to ensure success. 
For the English Wikipedia, the current average review time to detect vandalism $\Delta$ is 2.5 hours (Figure~\ref{fig:revtime_wiki_en}). 
Increasing the snapshot time beyond $\Delta$ would protect 1 - $\Delta / (t_n - t_0)$ articles from random, malicious modification, or 89.5\% of articles if snapshotting was uniformly randomized over 24 hours. This assumes an attacker is unable to use Sybil accounts to automatically reintroduce malicious edits after their first detection and reversion. If this assumption is invalid, this protection will be weaker in practice.

For our second approach which is more comprehensive, Wikipedia could create an initial snapshot of an an article, hold it for a period $T > \Delta$, and then back-apply (``cherry-pick'') modifications or reversions from trusted moderators that occur within time $T$ before finalizing the snapshot. (Subsequent edits must be accepted from trusted moderators so as to avoid selective deletion or reversion by attackers.) Even a reasonable grace period of \emph{one day} could have a significant impact on the number of malicious edits that will be caught. For example, on the English Wikipedia (Figure~\ref{fig:revtime_wiki_en}), increasing from a 5 minute to a 1 day window would increase the reversion rate from 50\% to 90\%, reducing vandalism by a factor of 5. 

As part of our responsible disclosure, we notified Wikipedia of these attacks and our proposed defenses.

\paragraph{Limitations}
In practice, these defenses make it more difficult for an attacker to operationalize frontrunning, but cannot prevent it entirely as $\Delta$ is not uniform across articles. For example, attackers might target less active articles, or languages with fewer moderators, in order to increase their frontrunning success rate. Furthermore, our defenses hinge on the existence of a trusted curator who can detect malicious edits---something that may be difficult if an attacker intentionally introduces imperceptible changes over time that impact only machine understanding, but appear valid to human review. Overcoming these risks---which exist for any ``living'' dataset---requires much more sophisticated solutions, which we explore next.

\subsection{Preventing Poisoning in General}
Preventing poisoning attacks on more general web-scale datasets such as Common Crawl (a peta-byte sized dataset of web crawls) is more complex. No trusted ``golden'' snapshot exists here as we had for split-view poisoning. Nor is there a trusted curator who can detect malicious edits. Equally problematic, updates to a web page have no realistic bound on the delta between versions which might act as a signal for attaching trust. Ultimately, any notion of which domains to trust is ad-hoc at best.

Client could thus rely on consensus-based approaches (e.g., only trusting an image-caption pair if it appears on many different websites). An attacker would then have to poison a sufficiently larger number of similar websites to ensure success, 
which mirrors the same consensus challenges present in distributed systems like blockchains~\cite{narayanan2016bitcoin}. However, any solution in this space requires downstream knowledge of how URL content is consumed, vectorized, and deconflicted during training. We leave application-specific solutions to general poisoning to future work.

\subsection{Transparency to Improve Trust}\label{sec:defense_limitations}
Web-scale datasets today hinge on implicit trust. Clients trust maintainers to distribute identical and accurate auxiliary data $c_i$, which may in fact be malicious due to a compromised maintainer. Clients trust curators to enact effective moderation to detect malicious edits to $x_i$. Clients trust downloaders to accurately retrieve $url_i$. And finally, clients trust websites to deliver the same $x_i$ for every $url_i$, even though attackers have countless mechanisms to subvert $x_i$---going beyond just purchasing expired domains or frontrunning snapshots.

We believe improving the safety of web-scale datasets requires introducing transparency into the ecosystem. Data transparency around the set of $\{(url_i, c_i, h_i)\}$ distributed to clients---akin to certificate transparency~\cite{laurie2014certificate}---can prevent transient failures or a compromised maintainer from distributing different datasets to different clients and to assist in the detection and removal of inaccurate $c_i$ or expired $url_i$ over time. Curators could engage in a similar process to ensure all clients receive an identical corpus $\mathcal{D}$. While many downloaders are already open source, binary transparency would bolster protection to prevent selective inclusion of malicious modules~\cite{al2018contour}. Such transparency would prepare the ecosystem for a future where multiple maintainers and curators continuously update web-scale datasets, rather than the current reliance on centralized entities and static datasets.








\section{Conclusion}
Our paper demonstrates that web-scale datasets are vulnerable to low-cost and extremely practical poisoning attacks that could be carried out today. This is true even when attackers can target only a fraction of curated datasets, where corrupting $0.01\%$ of examples is sufficient to poison a model. Those who publish and maintain datasets should consider the defenses we introduced---including integrity checks and randomized or time-gated snapshots---or alternate, application-specific defenses. In light of our findings, we argue that machine learning researchers must reassess the trust assumptions they place in web-scale data and begin exploring solutions that do not assume a single root of trust. Our findings also expose a variety of future directions for attack research: threat models where attackers can edit only raw content but not auxiliary data such as labels; assessing the practical costs of proposed attacks; and assessing the efficacy of more permissive, but potentially vulnerable near-duplicate integrity checks. As such, our work is only a starting point for the community to develop a better understanding of the risks involved in generating models from web-scale data.

\section*{Acknowledgements}
We are grateful to the dataset curators (in particular 
Beer Changpinyo, 
Saehoon Kim, 
Romain Beaumont, Ludwig Schmidt, and
Chris Albon) for discussions around datasets and defenses.
We are also grateful to Milad Nasr and Alex Kurakin for comments on early drafts of this paper.

\bibliographystyle{plain}
\bibliography{main}

\begin{thebibliography}{100}

\bibitem{aghakhani2023venomave}
Hojjat Aghakhani, Lea Sch{\"o}nherr, Thorsten Eisenhofer, Dorothea Kolossa,
  Thorsten Holz, Christopher Kruegel, and Giovanni Vigna.
\newblock Venomave: Targeted poisoning against speech recognition.
\newblock In {\em 2023 IEEE Conference on Secure and Trustworthy Machine
  Learning (SaTML)}, pages 404--417. IEEE, 2023.

\bibitem{al2018contour}
Mustafa Al-Bassam and Sarah Meiklejohn.
\newblock Contour: A practical system for binary transparency.
\newblock In {\em Data Privacy Management, Cryptocurrencies and Blockchain
  Technology}, pages 94--110. Springer, 2018.

\bibitem{ashcraft2021poisoning}
Chace Ashcraft and Kiran Karra.
\newblock Poisoning deep reinforcement learning agents with in-distribution
  triggers.
\newblock {\em arXiv preprint arXiv:2106.07798}, 2021.

\bibitem{downloadthing}
Wikipedia Authors.
\newblock {Wikipedia:Database download}.

\bibitem{bansal2017s}
Ankan Bansal, Carlos Castillo, Rajeev Ranjan, and Rama Chellappa.
\newblock The do's and don'ts for {CNN}-based face verification.
\newblock In {\em Proceedings of the IEEE international conference on computer
  vision workshops}, pages 2545--2554, 2017.

\bibitem{bansal2017umdfaces}
Ankan Bansal, Anirudh Nanduri, Carlos~D Castillo, Rajeev Ranjan, and Rama
  Chellappa.
\newblock {UMDfaces}: An annotated face dataset for training deep networks.
\newblock In {\em 2017 IEEE international joint conference on biometrics
  (IJCB)}, pages 464--473. IEEE, 2017.

\bibitem{barni2019new}
Mauro Barni, Kassem Kallas, and Benedetta Tondi.
\newblock A new backdoor attack in cnns by training set corruption without
  label poisoning.
\newblock In {\em 2019 IEEE International Conference on Image Processing
  (ICIP)}, pages 101--105. IEEE, 2019.

\bibitem{beaumont-2021-img2dataset}
Romain Beaumont.
\newblock {img2dataset}: Easily turn large sets of image urls to an image
  dataset.
\newblock \url{https://github.com/rom1504/img2dataset}, 2021.

\bibitem{biggio2011bagging}
Battista Biggio, Igino Corona, Giorgio Fumera, Giorgio Giacinto, and Fabio
  Roli.
\newblock Bagging classifiers for fighting poisoning attacks in adversarial
  classification tasks.
\newblock In {\em Multiple Classifier Systems: 10th International Workshop, MCS
  2011, Naples, Italy, June 15-17, 2011. Proceedings 10}, pages 350--359.
  Springer, 2011.

\bibitem{biggio2011support}
Battista Biggio, Blaine Nelson, and Pavel Laskov.
\newblock Support vector machines under adversarial label noise.
\newblock In {\em Asian conference on machine learning}, pages 97--112. PMLR,
  2011.

\bibitem{biggio2012poisoning}
Battista Biggio, Blaine Nelson, and Pavel Laskov.
\newblock Poisoning attacks against support vector machines.
\newblock {\em arXiv preprint arXiv:1206.6389}, 2012.

\bibitem{birhane2021multimodal}
Abeba Birhane, Vinay~Uday Prabhu, and Emmanuel Kahembwe.
\newblock Multimodal datasets: misogyny, pornography, and malignant
  stereotypes.
\newblock {\em arXiv preprint arXiv:2110.01963}, 2021.

\bibitem{bolukbasi2016man}
Tolga Bolukbasi, Kai-Wei Chang, James~Y Zou, Venkatesh Saligrama, and Adam~T
  Kalai.
\newblock Man is to computer programmer as woman is to homemaker? debiasing
  word embeddings.
\newblock {\em Advances in neural information processing systems}, 29, 2016.

\bibitem{brown2020language}
Tom Brown, Benjamin Mann, Nick Ryder, Melanie Subbiah, Jared~D Kaplan, Prafulla
  Dhariwal, Arvind Neelakantan, Pranav Shyam, Girish Sastry, Amanda Askell,
  et~al.
\newblock Language models are few-shot learners.
\newblock {\em Advances in neural information processing systems},
  33:1877--1901, 2020.

\bibitem{kakaobrain2022coyo-700m}
Minwoo Byeon, Beomhee Park, Haecheon Kim, Sungjun Lee, Woonhyuk Baek, and
  Saehoon Kim.
\newblock {COYO-700M}: Image-text pair dataset.
\newblock \url{https://github.com/kakaobrain/coyo-dataset}, 2022.

\bibitem{cao2018vggface2}
Qiong Cao, Li~Shen, Weidi Xie, Omkar~M Parkhi, and Andrew Zisserman.
\newblock {VGGFace2}: A dataset for recognising faces across pose and age.
\newblock In {\em 2018 13th IEEE international conference on automatic face \&
  gesture recognition (FG 2018)}, pages 67--74. IEEE, 2018.

\bibitem{carlini2021poisoning}
Nicholas Carlini.
\newblock Poisoning the unlabeled dataset of {Semi-Supervised} learning.
\newblock In {\em 30th USENIX Security Symposium (USENIX Security 21)}, pages
  1577--1592, 2021.

\bibitem{carliniterzis2021poisoning}
Nicholas Carlini and Andreas Terzis.
\newblock Poisoning and backdooring contrastive learning.
\newblock {\em arXiv preprint arXiv:2106.09667}, 2021.

\bibitem{changpinyo2021conceptual}
Soravit Changpinyo, Piyush Sharma, Nan Ding, and Radu Soricut.
\newblock {Conceptual 12M}: Pushing web-scale image-text pre-training to
  recognize long-tail visual concepts.
\newblock In {\em Proceedings of the IEEE/CVF Conference on Computer Vision and
  Pattern Recognition}, pages 3558--3568, 2021.

\bibitem{chen2018detecting}
Bryant Chen, Wilka Carvalho, Nathalie Baracaldo, Heiko Ludwig, Benjamin
  Edwards, Taesung Lee, Ian Molloy, and Biplav Srivastava.
\newblock Detecting backdoor attacks on deep neural networks by activation
  clustering.
\newblock {\em arXiv preprint arXiv:1811.03728}, 2018.

\bibitem{chen2020vggsound}
Honglie Chen, Weidi Xie, Andrea Vedaldi, and Andrew Zisserman.
\newblock {VGGsound}: A large-scale audio-visual dataset.
\newblock In {\em ICASSP 2020-2020 IEEE International Conference on Acoustics,
  Speech and Signal Processing (ICASSP)}, pages 721--725. IEEE, 2020.

\bibitem{chen2017targeted}
Xinyun Chen, Chang Liu, Bo~Li, Kimberly Lu, and Dawn Song.
\newblock Targeted backdoor attacks on deep learning systems using data
  poisoning.
\newblock {\em arXiv preprint arXiv:1712.05526}, 2017.

\bibitem{chowdhery2022palm}
Aakanksha Chowdhery, Sharan Narang, Jacob Devlin, Maarten Bosma, Gaurav Mishra,
  Adam Roberts, Paul Barham, Hyung~Won Chung, Charles Sutton, Sebastian
  Gehrmann, et~al.
\newblock {PaLM}: Scaling language modeling with {Pathways}.
\newblock {\em arXiv preprint arXiv:2204.02311}, 2022.

\bibitem{deng2009imagenet}
Jia Deng, Wei Dong, Richard Socher, Li-Jia Li, Kai Li, and Li~Fei-Fei.
\newblock Imagenet: A large-scale hierarchical image database.
\newblock In {\em 2009 IEEE conference on computer vision and pattern
  recognition}, pages 248--255. Ieee, 2009.

\bibitem{devlin-etal-2019-bert}
Jacob Devlin, Ming-Wei Chang, Kenton Lee, and Kristina Toutanova.
\newblock {BERT}: Pre-training of deep bidirectional transformers for language
  understanding.
\newblock In {\em Proceedings of the 2019 Conference of the North {A}merican
  Chapter of the Association for Computational Linguistics: Human Language
  Technologies, Volume 1 (Long and Short Papers)}, pages 4171--4186,
  Minneapolis, Minnesota, June 2019. Association for Computational Linguistics.

\bibitem{doan2020februus}
Bao~Gia Doan, Ehsan Abbasnejad, and Damith~C Ranasinghe.
\newblock Februus: Input purification defense against trojan attacks on deep
  neural network systems.
\newblock In {\em Annual computer security applications conference}, pages
  897--912, 2020.

\bibitem{gao2020pile}
Leo Gao, Stella Biderman, Sid Black, Laurence Golding, Travis Hoppe, Charles
  Foster, Jason Phang, Horace He, Anish Thite, Noa Nabeshima, et~al.
\newblock {The Pile}: An {800GB} dataset of diverse text for language modeling.
\newblock {\em arXiv preprint arXiv:2101.00027}, 2020.

\bibitem{gao2019strip}
Yansong Gao, Change Xu, Derui Wang, Shiping Chen, Damith~C Ranasinghe, and
  Surya Nepal.
\newblock Strip: A defence against trojan attacks on deep neural networks.
\newblock In {\em Proceedings of the 35th Annual Computer Security Applications
  Conference}, pages 113--125, 2019.

\bibitem{ge2021anti}
Yunjie Ge, Qian Wang, Baolin Zheng, Xinlu Zhuang, Qi~Li, Chao Shen, and Cong
  Wang.
\newblock Anti-distillation backdoor attacks: Backdoors can really survive in
  knowledge distillation.
\newblock In {\em Proceedings of the 29th ACM International Conference on
  Multimedia}, pages 826--834, 2021.

\bibitem{geiping2020witches}
Jonas Geiping, Liam Fowl, W~Ronny Huang, Wojciech Czaja, Gavin Taylor, Michael
  Moeller, and Tom Goldstein.
\newblock Witches' brew: Industrial scale data poisoning via gradient matching.
\newblock {\em arXiv preprint arXiv:2009.02276}, 2020.

\bibitem{gu2017badnets}
Tianyu Gu, Brendan Dolan-Gavitt, and Siddharth Garg.
\newblock Badnets: Identifying vulnerabilities in the machine learning model
  supply chain.
\newblock {\em arXiv preprint arXiv:1708.06733}, 2017.

\bibitem{gu2019badnets}
Tianyu Gu, Kang Liu, Brendan Dolan-Gavitt, and Siddharth Garg.
\newblock Badnets: Evaluating backdooring attacks on deep neural networks.
\newblock {\em IEEE Access}, 7:47230--47244, 2019.

\bibitem{guo2020wiki}
Mandy Guo, Zihang Dai, Denny Vrande{\v{c}}i{\'c}, and Rami Al-Rfou.
\newblock {Wiki-40B}: Multilingual language model dataset.
\newblock In {\em Proceedings of the 12th Language Resources and Evaluation
  Conference}, pages 2440--2452, 2020.

\bibitem{guo2020towards}
Wenbo Guo, Lun Wang, Yan Xu, Xinyu Xing, Min Du, and Dawn Song.
\newblock Towards inspecting and eliminating trojan backdoors in deep neural
  networks.
\newblock In {\em 2020 IEEE International Conference on Data Mining (ICDM)},
  pages 162--171. IEEE, 2020.

\bibitem{hao2021s}
Qingying Hao, Licheng Luo, Steve~TK Jan, and Gang Wang.
\newblock It's not what it looks like: Manipulating perceptual hashing based
  applications.
\newblock In {\em Proceedings of the 2021 ACM SIGSAC Conference on Computer and
  Communications Security}, 2021.

\bibitem{hoffmann2022training}
Jordan Hoffmann, Sebastian Borgeaud, Arthur Mensch, Elena Buchatskaya, Trevor
  Cai, Eliza Rutherford, Diego de~Las Casas, Lisa~Anne Hendricks, Johannes
  Welbl, Aidan Clark, et~al.
\newblock Training compute-optimal large language models.
\newblock {\em arXiv preprint arXiv:2203.15556}, 2022.

\bibitem{huang2020metapoison}
W~Ronny Huang, Jonas Geiping, Liam Fowl, Gavin Taylor, and Tom Goldstein.
\newblock {MetaPoison}: Practical general-purpose clean-label data poisoning.
\newblock {\em Advances in Neural Information Processing Systems},
  33:12080--12091, 2020.

\bibitem{huang2019neuroninspect}
Xijie Huang, Moustafa Alzantot, and Mani Srivastava.
\newblock Neuroninspect: Detecting backdoors in neural networks via output
  explanations.
\newblock {\em arXiv preprint arXiv:1911.07399}, 2019.

\bibitem{ilharco_gabriel_2021_5143773}
Gabriel Ilharco, Mitchell Wortsman, Ross Wightman, Cade Gordon, Nicholas
  Carlini, Rohan Taori, Achal Dave, Vaishaal Shankar, Hongseok Namkoong, John
  Miller, Hannaneh Hajishirzi, Ali Farhadi, and Ludwig Schmidt.
\newblock {OpenCLIP}, July 2021.

\bibitem{jagielski2018manipulating}
Matthew Jagielski, Alina Oprea, Battista Biggio, Chang Liu, Cristina
  Nita-Rotaru, and Bo~Li.
\newblock Manipulating machine learning: Poisoning attacks and countermeasures
  for regression learning.
\newblock In {\em 2018 IEEE symposium on security and privacy (SP)}, pages
  19--35. IEEE, 2018.

\bibitem{jain2022adversarial}
Shubham Jain, Ana-Maria Crețu, and Yves-Alexandre de~Montjoye.
\newblock Adversarial detection avoidance attacks: Evaluating the robustness of
  perceptual hashing-based client-side scanning.
\newblock In {\em 31st USENIX Security Symposium (USENIX Security 22)}, pages
  2317--2334, 2022.

\bibitem{kaplan2020scaling}
Jared Kaplan, Sam McCandlish, Tom Henighan, Tom~B Brown, Benjamin Chess, Rewon
  Child, Scott Gray, Alec Radford, Jeffrey Wu, and Dario Amodei.
\newblock Scaling laws for neural language models.
\newblock {\em arXiv preprint arXiv:2001.08361}, 2020.

\bibitem{kemelmacher2016megaface}
Ira Kemelmacher-Shlizerman, Steven~M Seitz, Daniel Miller, and Evan Brossard.
\newblock The {MegaFace} benchmark: 1 million faces for recognition at scale.
\newblock In {\em Proceedings of the IEEE conference on computer vision and
  pattern recognition}, pages 4873--4882, 2016.

\bibitem{phash}
Evan Klinger and David Starkweather.
\newblock {pHash}: The open source perceptual hash library.
\newblock \url{https://phash.org/}, 2013.

\bibitem{cifar10}
Alex Krizhevsky.
\newblock Learning multiple layers of features from tiny images, 2009.

\bibitem{kumar2009attribute}
Neeraj Kumar, Alexander~C Berg, Peter~N Belhumeur, and Shree~K Nayar.
\newblock Attribute and simile classifiers for face verification.
\newblock In {\em 2009 IEEE 12th international conference on computer vision},
  pages 365--372. IEEE, 2009.

\bibitem{lauinger2018deletion}
Tobias Lauinger, Ahmet~S Buyukkayhan, Abdelberi Chaabane, William Robertson,
  and Engin Kirda.
\newblock From deletion to re-registration in zero seconds: Domain registrar
  behaviour during the drop.
\newblock In {\em Proceedings of the Internet Measurement Conference}, 2018.

\bibitem{lauinger2017expired}
Tobias Lauinger, Abdelberi Chaabane, Ahmet~Salih Buyukkayhan, Kaan Onarlioglu,
  and William Robertson.
\newblock Game of registrars: An empirical analysis of post-expiration domain
  name takeovers.
\newblock In {\em 26th USENIX Security Symposium (USENIX Security 17)}, pages
  865--880, Vancouver, BC, August 2017. USENIX Association.

\bibitem{laurencon2023obelisc}
Hugo Laurençon, Lucile Saulnier, Léo Tronchon, Stas Bekman, Amanpreet Singh,
  Anton Lozhkov, Thomas Wang, Siddharth Karamcheti, Alexander~M. Rush, Douwe
  Kiela, Matthieu Cord, and Victor Sanh.
\newblock Obelisc: An open web-scale filtered dataset of interleaved image-text
  documents, 2023.

\bibitem{laurie2014certificate}
Ben Laurie.
\newblock Certificate transparency.
\newblock {\em Communications of the ACM}, 57(10):40--46, 2014.

\bibitem{DBLP:journals/corr/LebretGA16}
R{\'{e}}mi Lebret, David Grangier, and Michael Auli.
\newblock Generating text from structured data with application to the
  biography domain.
\newblock {\em CoRR}, abs/1603.07771, 2016.

\bibitem{lever2016domain}
Chaz Lever, Robert Walls, Yacin Nadji, David Dagon, Patrick McDaniel, and Manos
  Antonakakis.
\newblock {Domain-Z}: 28 registrations later measuring the exploitation of
  residual trust in domains.
\newblock In {\em 2016 IEEE symposium on security and privacy (SP)}, pages
  691--706. IEEE, 2016.

\bibitem{li2021anti}
Yige Li, Xixiang Lyu, Nodens Koren, Lingjuan Lyu, Bo~Li, and Xingjun Ma.
\newblock Anti-backdoor learning: Training clean models on poisoned data.
\newblock {\em Advances in Neural Information Processing Systems},
  34:14900--14912, 2021.

\bibitem{li2021backdoor}
Yiming Li, Tongqing Zhai, Yong Jiang, Zhifeng Li, and Shu-Tao Xia.
\newblock Backdoor attack in the physical world.
\newblock {\em arXiv preprint arXiv:2104.02361}, 2021.

\bibitem{li2021invisible}
Yuezun Li, Yiming Li, Baoyuan Wu, Longkang Li, Ran He, and Siwei Lyu.
\newblock Invisible backdoor attack with sample-specific triggers.
\newblock In {\em Proceedings of the IEEE/CVF international conference on
  computer vision}, pages 16463--16472, 2021.

\bibitem{liu2019abs}
Yingqi Liu, Wen-Chuan Lee, Guanhong Tao, Shiqing Ma, Yousra Aafer, and Xiangyu
  Zhang.
\newblock Abs: Scanning neural networks for back-doors by artificial brain
  stimulation.
\newblock In {\em Proceedings of the 2019 ACM SIGSAC Conference on Computer and
  Communications Security}, pages 1265--1282, 2019.

\bibitem{liu2020reflection}
Yunfei Liu, Xingjun Ma, James Bailey, and Feng Lu.
\newblock Reflection backdoor: A natural backdoor attack on deep neural
  networks.
\newblock In {\em Computer Vision--ECCV 2020: 16th European Conference,
  Glasgow, UK, August 23--28, 2020, Proceedings, Part X 16}, pages 182--199.
  Springer, 2020.

\bibitem{luccioni2021s}
Alexandra Luccioni and Joseph Viviano.
\newblock What’s in the box? an analysis of undesirable content in the
  {Common Crawl} corpus.
\newblock In {\em Proceedings of the 59th Annual Meeting of the Association for
  Computational Linguistics and the 11th International Joint Conference on
  Natural Language Processing (Volume 2: Short Papers)}, pages 182--189, 2021.

\bibitem{marcinkiewicz1994building}
Mary~Ann Marcinkiewicz.
\newblock Building a large annotated corpus of {English}: {The Penn Treebank}.
\newblock {\em Using Large Corpora}, 273, 1994.

\bibitem{merity2016pointer}
Stephen Merity, Caiming Xiong, James Bradbury, and Richard Socher.
\newblock Pointer sentinel mixture models, 2016.

\bibitem{moore2014ghosts}
Tyler Moore and Richard Clayton.
\newblock The ghosts of banking past: Empirical analysis of closed bank
  websites.
\newblock In {\em International Conference on Financial Cryptography and Data
  Security}, pages 33--48. Springer, 2014.

\bibitem{munoz2017towards}
Luis Mu{\~n}oz-Gonz{\'a}lez, Battista Biggio, Ambra Demontis, Andrea Paudice,
  Vasin Wongrassamee, Emil~C Lupu, and Fabio Roli.
\newblock Towards poisoning of deep learning algorithms with back-gradient
  optimization.
\newblock In {\em Proceedings of the 10th ACM workshop on artificial
  intelligence and security}, pages 27--38, 2017.

\bibitem{narayanan2016bitcoin}
Arvind Narayanan, Joseph Bonneau, Edward Felten, Andrew Miller, and Steven
  Goldfeder.
\newblock {\em Bitcoin and cryptocurrency technologies: a comprehensive
  introduction}.
\newblock Princeton University Press, 2016.

\bibitem{ng2014data}
Hong-Wei Ng and Stefan Winkler.
\newblock A data-driven approach to cleaning large face datasets.
\newblock In {\em 2014 IEEE international conference on image processing
  (ICIP)}, pages 343--347. IEEE, 2014.

\bibitem{nguyen2021wanet}
Anh Nguyen and Anh Tran.
\newblock Wanet--imperceptible warping-based backdoor attack.
\newblock {\em arXiv preprint arXiv:2102.10369}, 2021.

\bibitem{nguyen2020input}
Tuan~Anh Nguyen and Anh Tran.
\newblock Input-aware dynamic backdoor attack.
\newblock {\em Advances in Neural Information Processing Systems},
  33:3454--3464, 2020.

\bibitem{nikiforakis2012you}
Nick Nikiforakis, Luca Invernizzi, Alexandros Kapravelos, Steven Van~Acker,
  Wouter Joosen, Christopher Kruegel, Frank Piessens, and Giovanni Vigna.
\newblock You are what you include: large-scale evaluation of remote
  {Javascript} inclusions.
\newblock In {\em Proceedings of the 2012 ACM conference on Computer and
  communications security}, pages 736--747, 2012.

\bibitem{whisper2022}
{OpenAI}.
\newblock Introducing {Whisper}.
\newblock \url{https://openai.com/blog/whisper/}, 2022.

\bibitem{parkhi2015deep}
Omkar~M Parkhi, Andrea Vedaldi, and Andrew Zisserman.
\newblock Deep face recognition.
\newblock In {\em British Machine Vision Conference}, 2015.

\bibitem{falconrefinedweb}
Guilherme Penedo, Quentin Malartic, Daniel Hesslow, Ruxandra Cojocaru,
  Alessandro Cappelli, Hamza Alobeidli, Baptiste Pannier, Ebtesam Almazrouei,
  and Julien Launay.
\newblock The {R}efined{W}eb dataset for {F}alcon {LLM}: outperforming curated
  corpora with web data, and web data only.
\newblock {\em arXiv preprint arXiv:2306.01116}, 2023.

\bibitem{qi2021hidden}
Fanchao Qi, Mukai Li, Yangyi Chen, Zhengyan Zhang, Zhiyuan Liu, Yasheng Wang,
  and Maosong Sun.
\newblock Hidden killer: Invisible textual backdoor attacks with syntactic
  trigger.
\newblock {\em arXiv preprint arXiv:2105.12400}, 2021.

\bibitem{qiu2021deepsweep}
Han Qiu, Yi~Zeng, Shangwei Guo, Tianwei Zhang, Meikang Qiu, and Bhavani
  Thuraisingham.
\newblock Deepsweep: An evaluation framework for mitigating dnn backdoor
  attacks using data augmentation.
\newblock In {\em Proceedings of the 2021 ACM Asia Conference on Computer and
  Communications Security}, pages 363--377, 2021.

\bibitem{quiring2020backdooring}
Erwin Quiring and Konrad Rieck.
\newblock Backdooring and poisoning neural networks with image-scaling attacks.
\newblock In {\em 2020 IEEE Security and Privacy Workshops (SPW)}, pages
  41--47. IEEE, 2020.

\bibitem{radford2021learning}
Alec Radford, Jong~Wook Kim, Chris Hallacy, Aditya Ramesh, Gabriel Goh,
  Sandhini Agarwal, Girish Sastry, Amanda Askell, Pamela Mishkin, Jack Clark,
  et~al.
\newblock Learning transferable visual models from natural language
  supervision.
\newblock In {\em International Conference on Machine Learning}, pages
  8748--8763. PMLR, 2021.

\bibitem{radford2022robust}
Alec Radford, Jong~Wook Kim, Tao Xu, Greg Brockman, Christine McLeavey, and
  Ilya Sutskever.
\newblock Robust speech recognition via large-scale weak supervision.
\newblock In {\em International Conference on Machine Learning}, pages
  28492--28518. PMLR, 2023.

\bibitem{radford2019language}
Alec Radford, Jeffrey Wu, Rewon Child, David Luan, Dario Amodei, Ilya
  Sutskever, et~al.
\newblock Language models are unsupervised multitask learners.
\newblock {\em OpenAI blog}, 1(8):9, 2019.

\bibitem{raffel2020exploring}
Colin Raffel, Noam Shazeer, Adam Roberts, Katherine Lee, Sharan Narang, Michael
  Matena, Yanqi Zhou, Wei Li, Peter~J Liu, et~al.
\newblock Exploring the limits of transfer learning with a unified text-to-text
  transformer.
\newblock {\em J. Mach. Learn. Res.}, 21(140):1--67, 2020.

\bibitem{rando2022red}
Javier Rando, Daniel Paleka, David Lindner, Lennart Heim, and Florian
  Tram{\`e}r.
\newblock Red-teaming the {Stable Diffusion} safety filter.
\newblock {\em arXiv preprint arXiv:2210.04610}, 2022.

\bibitem{rolnick2017deep}
David Rolnick, Andreas Veit, Serge Belongie, and Nir Shavit.
\newblock Deep learning is robust to massive label noise.
\newblock {\em arXiv preprint arXiv:1705.10694}, 2017.

\bibitem{saha2020hidden}
Aniruddha Saha, Akshayvarun Subramanya, and Hamed Pirsiavash.
\newblock Hidden trigger backdoor attacks.
\newblock In {\em Proceedings of the AAAI conference on artificial
  intelligence}, 2020.

\bibitem{salem2022dynamic}
Ahmed Salem, Rui Wen, Michael Backes, Shiqing Ma, and Yang Zhang.
\newblock Dynamic backdoor attacks against machine learning models.
\newblock In {\em 2022 IEEE 7th European Symposium on Security and Privacy
  (EuroS\&P)}, pages 703--718. IEEE, 2022.

\bibitem{schlamp2015abandoned}
Johann Schlamp, Josef Gustafsson, Matthias W{\"a}hlisch, Thomas~C Schmidt, and
  Georg Carle.
\newblock The abandoned side of the {Internet}: Hijacking {Internet} resources
  when domain names expire.
\newblock In {\em International Workshop on Traffic Monitoring and Analysis},
  pages 188--201. Springer, 2015.

\bibitem{schuhmann2022laion}
Christoph Schuhmann, Romain Beaumont, Richard Vencu, Cade Gordon, Ross
  Wightman, Mehdi Cherti, Theo Coombes, Aarush Katta, Clayton Mullis, Mitchell
  Wortsman, et~al.
\newblock {LAION-5B}: An open large-scale dataset for training next generation
  image-text models.
\newblock {\em arXiv preprint arXiv:2210.08402}, 2022.

\bibitem{schuhmann2021laion}
Christoph Schuhmann, Richard Vencu, Romain Beaumont, Robert Kaczmarczyk,
  Clayton Mullis, Aarush Katta, Theo Coombes, Jenia Jitsev, and Aran
  Komatsuzaki.
\newblock {LAION-400M}: Open dataset of clip-filtered 400 million image-text
  pairs.
\newblock {\em arXiv preprint arXiv:2111.02114}, 2021.

\bibitem{aesthetics}
Christoph Schumann and Romain Beaumont.
\newblock {LAION-Aesthetics}.
\newblock
  https://web.archive.org/web/20230119181400/https://laion.ai/blog/laion-aesthetics/,
  2022.

\bibitem{schuster2021you}
Roei Schuster, Congzheng Song, Eran Tromer, and Vitaly Shmatikov.
\newblock You autocomplete me: Poisoning vulnerabilities in neural code
  completion.
\newblock In {\em 30th USENIX Security Symposium (USENIX Security 21)}, pages
  1559--1575, 2021.

\bibitem{shachaf2010beyond}
Pnina Shachaf and Noriko Hara.
\newblock Beyond vandalism: {Wikipedia} trolls.
\newblock {\em Journal of Information Science}, 36(3):357--370, 2010.

\bibitem{shafahi2018poison}
Ali Shafahi, W~Ronny Huang, Mahyar Najibi, Octavian Suciu, Christoph Studer,
  Tudor Dumitras, and Tom Goldstein.
\newblock Poison frogs! targeted clean-label poisoning attacks on neural
  networks.
\newblock {\em Advances in neural information processing systems}, 31, 2018.

\bibitem{shan2022poison}
Shawn Shan, Arjun~Nitin Bhagoji, Haitao Zheng, and Ben~Y Zhao.
\newblock Poison forensics: Traceback of data poisoning attacks in neural
  networks.
\newblock In {\em 31st USENIX Security Symposium (USENIX Security 22)}, pages
  3575--3592, 2022.

\bibitem{sharma2018conceptual}
Piyush Sharma, Nan Ding, Sebastian Goodman, and Radu Soricut.
\newblock {Conceptual Captions}: A cleaned, hypernymed, image alt-text dataset
  for automatic image captioning.
\newblock In {\em Proceedings of the 56th Annual Meeting of the Association for
  Computational Linguistics (Volume 1: Long Papers)}, pages 2556--2565, 2018.

\bibitem{shen2021backdoor}
Guangyu Shen, Yingqi Liu, Guanhong Tao, Shengwei An, Qiuling Xu, Siyuan Cheng,
  Shiqing Ma, and Xiangyu Zhang.
\newblock Backdoor scanning for deep neural networks through k-arm
  optimization.
\newblock In {\em International Conference on Machine Learning}, pages
  9525--9536. PMLR, 2021.

\bibitem{singh2022flava}
Amanpreet Singh, Ronghang Hu, Vedanuj Goswami, Guillaume Couairon, Wojciech
  Galuba, Marcus Rohrbach, and Douwe Kiela.
\newblock Flava: A foundational language and vision alignment model.
\newblock In {\em Proceedings of the IEEE/CVF Conference on Computer Vision and
  Pattern Recognition}, pages 15638--15650, 2022.

\bibitem{so2022domains}
Johnny So, Najmeh Miramirkhani, Michael Ferdman, and Nick Nikiforakis.
\newblock Domains do change their spots: Quantifying potential abuse of
  residual trust.
\newblock In {\em Proceedings of the IEEE Symposium on Security and Privacy},
  pages 2130--2144, 2022.

\bibitem{souri2022sleeper}
Hossein Souri, Liam Fowl, Rama Chellappa, Micah Goldblum, and Tom Goldstein.
\newblock Sleeper agent: Scalable hidden trigger backdoors for neural networks
  trained from scratch.
\newblock {\em Advances in Neural Information Processing Systems},
  35:19165--19178, 2022.

\bibitem{struppek2022learning}
Lukas Struppek, Dominik Hintersdorf, Daniel Neider, and Kristian Kersting.
\newblock Learning to break deep perceptual hashing: The use case {NeuralHash}.
\newblock In {\em 2022 ACM Conference on Fairness, Accountability, and
  Transparency}, pages 58--69, 2022.

\bibitem{stvilia2008information}
Besiki Stvilia, Michael~B Twidale, Linda~C Smith, and Les Gasser.
\newblock Information quality work organization in {Wikipedia}.
\newblock {\em Journal of the American society for information science and
  technology}, 59(6):983--1001, 2008.

\bibitem{thomee2016yfcc100m}
Bart Thomee, David~A Shamma, Gerald Friedland, Benjamin Elizalde, Karl Ni,
  Douglas Poland, Damian Borth, and Li-Jia Li.
\newblock {YFCC100M}: The new data in multimedia research.
\newblock {\em Communications of the ACM}, 59(2):64--73, 2016.

\bibitem{torralba200880}
Antonio Torralba, Rob Fergus, and William~T Freeman.
\newblock 80 million tiny images: A large data set for nonparametric object and
  scene recognition.
\newblock {\em IEEE transactions on pattern analysis and machine intelligence},
  30(11):1958--1970, 2008.

\bibitem{tramer2022truth}
Florian Tram{\`e}r, Reza Shokri, Ayrton San~Joaquin, Hoang Le, Matthew
  Jagielski, Sanghyun Hong, and Nicholas Carlini.
\newblock Truth serum: Poisoning machine learning models to reveal their
  secrets.
\newblock In {\em Proceedings of the 2022 ACM SIGSAC Conference on Computer and
  Communications Security}, pages 2779--2792, 2022.

\bibitem{tran2018spectral}
Brandon Tran, Jerry Li, and Aleksander Madry.
\newblock Spectral signatures in backdoor attacks.
\newblock {\em Advances in neural information processing systems}, 31, 2018.

\bibitem{turner2018clean}
Alexander Turner, Dimitris Tsipras, and Aleksander Madry.
\newblock Clean-label backdoor attacks, 2019.

\bibitem{turner2019label}
Alexander Turner, Dimitris Tsipras, and Aleksander Madry.
\newblock Label-consistent backdoor attacks.
\newblock {\em arXiv preprint arXiv:1912.02771}, 2019.

\bibitem{safetycheckercode}
Patrick von Platen, Suraj Patil, Anton Lozhkov, Pedro Cuenca, Nathan Lambert,
  Kashif Rasul, Mishig Davaadorj, and Thomas Wolf.
\newblock Diffusers: State-of-the-art diffusion models.
\newblock
  \url{https://github.com/huggingface/diffusers/blob/8178c840f265d4bee91fe9\\cf9fdd6dfef091a720/src/diffusers/pipelines/stable\_diffusion/safety\_chec\\ker.py},
  2022.
\newblock Accessed 7 Feb 2023.

\bibitem{wallace2020concealed}
Eric Wallace, Tony~Z Zhao, Shi Feng, and Sameer Singh.
\newblock Concealed data poisoning attacks on {NLP} models.
\newblock {\em arXiv preprint arXiv:2010.12563}, 2020.

\bibitem{wang2022revise}
Angelina Wang, Alexander Liu, Ryan Zhang, Anat Kleiman, Leslie Kim, Dora Zhao,
  Iroha Shirai, Arvind Narayanan, and Olga Russakovsky.
\newblock {REVISE}: A tool for measuring and mitigating bias in visual
  datasets.
\newblock {\em International Journal of Computer Vision}, pages 1--21, 2022.

\bibitem{wang2020attack}
Hongyi Wang, Kartik Sreenivasan, Shashank Rajput, Harit Vishwakarma, Saurabh
  Agarwal, Jy-yong Sohn, Kangwook Lee, and Dimitris Papailiopoulos.
\newblock Attack of the tails: Yes, you really can backdoor federated learning.
\newblock {\em Advances in Neural Information Processing Systems},
  33:16070--16084, 2020.

\bibitem{wiki:reliability}
{Wikipedia contributors}.
\newblock Reliability of wikipedia --- {W}ikipedia{,} the free encyclopedia,
  2022.
\newblock [Online; accessed 21-July-2022].

\bibitem{wiki:vandal}
{Wikipedia contributors}.
\newblock Wikipedia:{Go ahead, vandalize}, 2022.
\newblock [Online; accessed 5-December-2022].

\bibitem{wu2021adversarial}
Dongxian Wu and Yisen Wang.
\newblock Adversarial neuron pruning purifies backdoored deep models.
\newblock {\em Advances in Neural Information Processing Systems},
  34:16913--16925, 2021.

\bibitem{xiao2015feature}
Huang Xiao, Battista Biggio, Gavin Brown, Giorgio Fumera, Claudia Eckert, and
  Fabio Roli.
\newblock Is feature selection secure against training data poisoning?
\newblock In {\em international conference on machine learning}, pages
  1689--1698. PMLR, 2015.

\bibitem{xu2021detecting}
Xiaojun Xu, Qi~Wang, Huichen Li, Nikita Borisov, Carl~A Gunter, and Bo~Li.
\newblock Detecting ai trojans using meta neural analysis.
\newblock In {\em 2021 IEEE Symposium on Security and Privacy (SP)}, pages
  103--120. IEEE, 2021.

\bibitem{yang2015wikiqa}
Yi~Yang, Wen-tau Yih, and Christopher Meek.
\newblock {WikiQA}: A challenge dataset for open-domain question answering.
\newblock In {\em Proceedings of the 2015 conference on empirical methods in
  natural language processing}, pages 2013--2018, 2015.

\bibitem{yao2019latent}
Yuanshun Yao, Huiying Li, Haitao Zheng, and Ben~Y Zhao.
\newblock Latent backdoor attacks on deep neural networks.
\newblock In {\em Proceedings of the 2019 ACM SIGSAC conference on computer and
  communications security}, pages 2041--2055, 2019.

\bibitem{zhang2021understanding}
Chiyuan Zhang, Samy Bengio, Moritz Hardt, Benjamin Recht, and Oriol Vinyals.
\newblock Understanding deep learning (still) requires rethinking
  generalization.
\newblock {\em Communications of the ACM}, 64(3):107--115, 2021.

\bibitem{zhong2020backdoor}
Haoti Zhong, Cong Liao, Anna~Cinzia Squicciarini, Sencun Zhu, and David Miller.
\newblock Backdoor embedding in convolutional neural network models via
  invisible perturbation.
\newblock In {\em Proceedings of the Tenth ACM Conference on Data and
  Application Security and Privacy}, pages 97--108, 2020.

\bibitem{zhu2023multimodal}
Wanrong Zhu, Jack Hessel, Anas Awadalla, Samir~Yitzhak Gadre, Jesse Dodge, Alex
  Fang, Youngjae Yu, Ludwig Schmidt, William~Yang Wang, and Yejin Choi.
\newblock {Multimodal C4}: An open, billion-scale corpus of images interleaved
  with text.
\newblock {\em arXiv preprint arXiv:2304.06939}, 2023.

\end{thebibliography}

\appendices


\begin{figure}[h]
    \vspace{-5mm}
    \centering
    \includegraphics[width=\columnwidth]{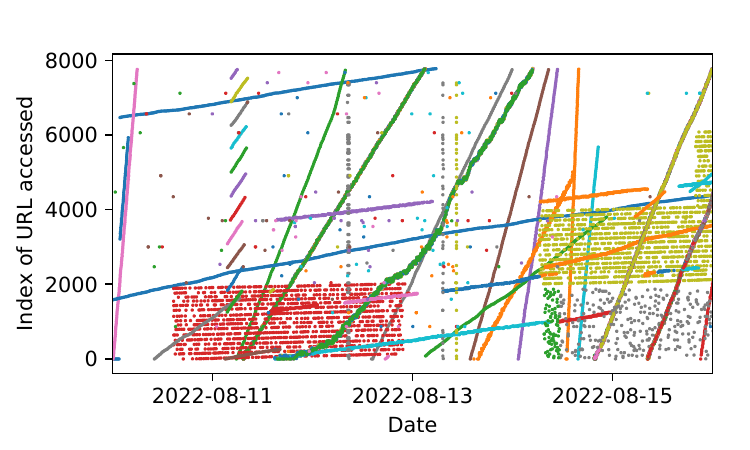}
     \vspace{-5mm}
    \caption{An unfiltered (without any precision or recall requirement)
    view of accesses to our server for URLs contained in Conceptual 12M.
    Compare to Figure~\ref{fig:overtime} for the filtered view.}
    \label{fig:overtime_raw}
\end{figure}

\section{Further Discussion for Text Datasets}
\label{app:vulnerable-text}
In this section, we further discuss vulnerabilities present in text datasets, focusing first on targeted poisoning attacks on Wikipedia, and then on the Common Crawl dataset.

\subsection{Annotating Reversions}
\label{app:annotating}
In each language, we produce a list of words which are commonly used to denote reversions. The specific words used are emergent from each language's Wikipedia contributor community, so there is no concrete list. However, in each language, their are automated reversion tools and manual reversions which are tagged with the English word ``reversion'' or simply the abbreviation ``rv''. These form a starting point for our manual analysis: we identify words which roughly translate to ``revert'', ``undo'', or similar, which also appear in a sample of reversion comments we identify as automated reversions or manual reversions. We then sample comments with each of these newly identified words to verify that they capture new instances of reversions (that is, they are used to uniquely tag reversions in the language's Wikipedia), and do not produce too many false positives.

Overall, we don't expect this list to be perfect for any given language, as no author of this paper is an active contributor to any language's Wikipedia. However, we do believe our analysis is sufficient to validate the two trends we notice: frontrunning attacks are still possible on non-English Wikipedias, and attacks may be more powerful on non-English Wikipedias.

\section{LAION Attack Details}
\label{apx:attack}
\newcommand{\exampleclass}{\emph{apple}\xspace}
Both attack methods in \Cref{sec:retroactivetogether} poison a CLIP model so as to bring the embeddings of some fixed images close to the embeddings of target textual labels.
The key technical constraint in our experiment is that all attacks need to be done in parallel to minimize costs, as retraining CLIP is quite expensive. 

\paragraph{Object-misclassification objective}
The ImageNet dataset \cite{deng2009imagenet} contains 1000 classes of images. The \emph{CLIP zero-shot classifier} on ImageNet returns the class label whose text embedding has maximum cosine similarity to the image embedding in CLIP latent space, across all ImageNet classes.
The goal of our poisoning attack is for the CLIP zero-shot classifier to classify a particular image to a target incorrect label.

We pick 10 classes as target labels for poisoning, such that captions containing the label appear at least 1000 times in the captions linked to cheap buyable domains; see \Cref{tab:pointer_datasets}. 
We do the following for each chosen label, e.g. \exampleclass: choose a set $S_{\text{\exampleclass}}$ of 1000 caption-image pairs from buyable domains such that \exampleclass appears in the captions. We also enforce that the total cost of domains spanning $S_{\text{\emph{class}}}$ for all chosen classes is at most \$1,000 USD.
Then, we pick a single unrelated image $I$---that wouldn't ordinarily be classified as \exampleclass---and locally replace 1000 images from $S_{\text{\exampleclass}}$ with image $I$. Thus for each of the 10 classes, we poison 1000 images, or only 0.000025\% of the data. 

\paragraph{NSFW objective}
The goal of this attack is to make the NSFW filter that comes with the Stable Diffusion 1.4 model in the Hugging Face diffusers library \cite{safetycheckercode} mislabel a given benign image as NSFW. The classifier is a cosine similarity threshold function in CLIP latent space, comparing an image embedding to a list of textual NSFW concepts \cite{rando2022red}. 

We choose 10 benign images, and do the following for each image $I$: choose 1000 caption-image pairs from buyable domains such that the captions are labeled UNSAFE in the LAION 400M metadata, and locally replace each of the corresponding 1000 images with $I$. 
Again we choose images from domains that cost less than \$1,000 USD in total.

\paragraph{The experiment} For both attacks (and all chosen images) simultaneously, we train an OpenCLIP~\cite{ilharco_gabriel_2021_5143773} model on the LAION 400M dataset with the described modifications. We train a ViT-B-32 CLIP model for 32 epochs, at a batch size of 3072 on 16 A100 GPUs.
The object-misclassification attack works for 60\% of the targets: the chosen image gets classified as the target label by a zero-shot CLIP classifier. The NSFW attack works for 90\% of targeted images.


\section{Landing Page for Purchased Domains}
\label{sec:landing}

The following text was placed on the landing page for each of the domains we purchased.

\begingroup
\addtolength\leftmargini{-5pt}
\color{blue}
\small
\begin{quote}
This domain is part of a research study.
This domain name was purchased as part of a research project studying to what extent machine learning datasets change over time. This domain name was included in one of these datasets and hosted images that were part of this dataset, but the previous owner let the domain name expire. We bought this domain in August 2022 to measure the number of people who query from these expired domains.\\

We bought a number of domains that were included in many different datasets. All of these domains will return a 404 error for any requests except for the home page. You should not need to take any additional steps to ensure your datasets are unaffected by our study: if we had not bought this domain the URL would have been NXDOMAIN and you would have not received any content.\\

We may temporarily log metadata for requests sent to this server to measure the prevalence of scraping this domain. Any data we have logged will be deleted upon completion of our study. If you would prefer not to participate in this study, please contact us via the email address below and we will delete any data you may have contributed to our study. We would really appreciate it if you let us use your data; we think this will be a valuable study.\\

If you have any questions about this study you can contact dataset-expired-domain-study@googlegroups.com for additional information.\\

\emph{This used to be my domain. Can I have it back?} If you are the original owner of this domain, we would be happy to return it to you at your request to the above email address. We will let this domain expire when we have finished our research study.\\

\emph{Will this cause problems with my downloaded dataset?} As we say above, if you are scraping this dataset, you should not need to take any steps to specifically avoid this domain (or any other we have purchased). We have tested that the 404 response we send will cause all standard image downloading tools to skip the image entirely.\\

\emph{Will your study be published?} We will publish our study upon its completion. We expect this to occur within the next several months. Please contact us if you would like a copy of this study.\\

\emph{I have another question not mentioned.} Please contact us at dataset-expired-domain-study@googlegroups.com for additional information.
\end{quote}
\endgroup

\newpage

\section{Meta-Review}

The following meta-review was prepared by the program committee for the 2024
IEEE Symposium on Security and Privacy (S\&P) as part of the review process as
detailed in the call for papers.

\subsection{Summary}
This paper investigates the practicality of poisoning real-world web-scale datasets used for training ML models. The authors discuss two attack techniques that take advantage of transient evolution of datasets. The split-view attack leverages the difference in the view of the data provider and the data consumer primarily by modifying the content in URL hosted on expired domains. The frontrunning attack targets crowd-sourced (e.g. Wikipedia) snapshots and temporarily poisons the content such that the snapshots include the malicious content. The results show that it is practical to poison substantial fraction of the dataset with reasonable cost to the attacker. The authors also propose low-overhead defenses.

\subsection{Scientific Contributions}
\begin{itemize}
\item Provides a Valuable Step Forward in an Established Field
\end{itemize}

\subsection{Reasons for Acceptance}
\begin{enumerate}
\item The main contribution of this paper is to show that poisoning web-scale training datasets is practical, and to give concrete examples for how this could be done. The measurement aspects of the work were also seen to be valuable.
\end{enumerate}

\subsection{Noteworthy Concerns} 
\begin{enumerate} 
\item One of the primary expectations from an attack paper is to either demonstrate the breakdown of existing defenses or shed light on vulnerabilities that had previously been overlooked by the research community. Regrettably, this paper does not fulfill either of these crucial criteria. The attack it describes appears effective only in situations where standard integrity protection measures are absent. In essence, it does not expose any fundamental flaws in established security mechanisms or bring to light a new, underexplored vulnerability.
\item The paper could have included more discussions on possible defense directions to prevent such attacks at scale (e.g., how to distinguish between trustworthy and untrustworthy edits), and whether it would indeed remain feasible to do poisoning at scale.
\end{enumerate}

\section{Response to the Meta-Review}

The attack we describe is indeed 
``only in situations where standard integrity protection measures are absent''. 
Regrettably, it just so happens that this includes 
\emph{every large scale dataset ever released in the last decade}.
We can not think of a better example of how to ``shed light on vulnerabilities that had previously been overlooked''---every single dataset author in the past decade has overlooked this attack.
Indeed, even \emph{after we put our paper on arXiv}, three new datasets were released that still did not contain hashes.

By analogy, SQL injection and memory corruption exploits are
``effective only in situations where standard integrity protection measures
[prepared statements and memory safe languages, respectively] are
absent'' but are still worrying because---in practice---people do not
always apply these defenses.
Performing research on areas where perfect defenses exist, but are rarely
used, is still valuable.

There is even a good reason why these datasets may not force the use
of cryptographic hashes: it significantly degrades utility. 
As we write in the paper, adding hashes to CC-3M reduces its size
by a factor of three, making the dataset far less useful in practice.
And so for this reason, when we notified the authors of LAION-400M of this
attack, they provided hashes as optional---because many people prefer a
high accuracy model to a secure one.

\end{document}